\def\nome#1{{ \label{#1} }}
\def\writenote#1{{ }}

\def\COMMENTA#1{{}}

\def\partder#1#2{ {\partial #1 \over \partial #2} }
\def\primato#1{{{#1}^\prime}}

\def\Be{\begin{equation}}
\def\Ba{\begin{array}}
\def\Ee{\end{equation}}
\def\Ea{\end{array}}

\def\Ds{\displaystyle }

\def\OldRefP#1{{}}                              

\def\half{{1\over2}}

\def\EL{\mbox{$\cal E\! L$}}
\def\deq{{\buildrel \rm \circ \over =}}

\def\BMx{{\mbox{\boldmath $x$}}}

\def\BMz{{\mbox{\boldmath $z$}}}

\def\Fc{{\cal F}}

\def\Hc{{\cal H}}

\def\Lc{{\cal L}}
\def\Mc{{\cal M}}
\def\Nc{{\cal N}}

\def\Sc{{\cal S}}

\documentstyle[12pt]{article}
\begin{document}

\addtolength{\baselineskip}{.5 \baselineskip}
\parskip = 2ex

\renewcommand{\theequation}{\thesection.\arabic{equation}}


\begin{flushright}
Short title: {\bf Generalized Newtonian Theories of Gravitation.} \\
P.A.C.S. number:+04.50,04.90,03.50
\end{flushright}
\begin{center}
{ \LARGE \bf STANDARD AND GENERALIZED \\
    NEWTONIAN GRAVITIES AS \\
    ``GAUGE'' THEORIES OF \\ 
    THE EXTENDED GALILEI GROUP - \\[2 mm]
     II: DYNAMICAL THREE-SPACE THEORIES}
\end{center}
\begin{center}
        R.~DE PIETRI${}^1$  \\[1 mm]
{\it Department of  Physics and Astronomy, University of Pittsburgh, } \\
{\it Pittsburgh, PA 15260, USA }\\
\vskip 2 mm
L.~LUSANNA  \\[1 mm]
{\it I.N.F.N., Sezione di Firenze}  \\
{\it Largo E. Fermi 2, } {\it 50127 Arcetri (FI), Italy} \\
\vskip 1mm
and    \\
\vskip 1mm
        M. PAURI\footnote{On leave from: 
{Dipartimento di Fisica - Sezione di Fisica Teorica},
{Universit\`a di Parma, 43100 Parma, Italy,  } 
{and   I.N.F.N., Sezione di Milano, Gruppo Collegato di Parma}}
 \\[1 mm]
{\it Center for Philosophy of Science - University of Pittsburgh} \\
{\it 817D Cathedral of Learning, Pittsburgh, PA 15260, USA }\\
\end{center}
\newpage
\begin{center} {\large \bf Abstract} \end{center}
\begin{quote}
In a preceding paper we developed a reformulation of Newtonian
gravitation as a {\it gauge} theory of the extended Galilei group.
In the present one we derive two 
true generalizations of Newton's theory (a {\it ten-fields} and 
an {\it eleven-fields} theory),
in terms of an explicit Lagrangian realization of the 
{\it absolute time} dynamics of a Riemannian three-space.
They turn out to be {\it gauge invariant} theories of the extended
Galilei group in the same sense in which general relativity is said
to be a {\it gauge} theory of the Poincar\'e group. The {\it ten-fields} 
theory provides a dynamical realization of some of the so-called
``Newtonian space-time structures'' which have been
geometrically classified by K\"{u}nzle and Kucha\v{r}.
The {\it eleven-fields} theory involves a {\it dilaton-like} scalar
potential in addition to Newton's potential and, like general relativity,
has a three-metric with {\it two} dynamical degrees of freedom.
It is interesting to find that, 
within the linear approximation, such degrees of freedom
show {\it
graviton-like} features: they satisfy a wave equation and
propagate with a velocity related to the scalar
Newtonian potential.
\end{quote}

\newpage
\setcounter{equation}{0}
\section{Introduction}

        The present paper must be seen, from the technical point 
of view, as a natural development of a previous one \cite{DePietri},
in which Newtonian gravitation has been reformulated as a {\it gauge}
theory of the {\it extended} Galilei group. The role of the Galilei 
group has thereby been transformed from that of a 
{\it symmetry} group to that of a {\it covariance} group, in the sense
of Anderson \cite{ANDE}. Yet, from a substantial point of view, this
paper contains true generalizations of Newton's theory: we propose here
two new gravitational theories which provide non-relativistic (Galilean)
Lagrangian descriptions of the evolution 
of a Riemannian dynamical three-space in {\it absolute time}.

      The recasting of Newton's theory has been obtained in \cite{DePietri}
in the form of a {\it three-dimensional Galilean} ({\it absolute time} 
respecting) {\it generally-covariant} Action principle. 
 Thus our three-dimensional
formulation differs
substantially from the well-known four-dimensional reformulations of 
Newton's theory in geometrical terms
by \'Elie Cartan \cite{Cartan}, Havas \cite{Havas}, Anderson \cite{ANDE},
Trautman \cite{TrautA}, K\"unzle \cite{Kunz} and Kucha\v{r} \cite{Kuch}.
All these reformulations describe the Newtonian 
inertial-gravitational structure in terms of an affine connection 
compatible with the temporal flow $t_{\mu}$ and a rank-three spatial
metric $h^{\mu\nu}$. While the curvature of the four-dimensional affine
connection is different from zero because of the presence of matter, the
Newtonian flatness of the absolute three-space is guaranteed by the further
requirement that Poisson's equation be satisfied, in the covariant
form $R_{\mu\nu} = G \rho (z) t_\mu t_\nu$, where $R_{\mu\nu}$ is the Ricci 
tensor of the affine connection and $\rho(z)$ is the matter density. In 
this way, the four-dimensional description is {\it dynamical}, while the
three-dimensional one is not. 

It is clear that, in order to achieve the main
scope we are interested in, it necessary first of all 
to get rid, in some way, of the flatness 
condition of the
absolute three-space metric $g_{ij}$, which is expressed by the 
above covariant 
Poisson equation in the four-dimensional formulations and 
by the explicit vanishing of the three-dimensional Ricci
tensor ($R_{ij} = 0$) in our three-dimensional formulation.
Now, in \cite{DePietri} we have obtained essentially the following results:
(1) By exploiting the {\it gauge} methodology originally applied by Utiyama
\cite{Uty} to the Lorentz group within 
the field theoretic framework, all the inertial-gravitational fields 
which can be coupled to a non-relativistic
mass-point have been characterized.
(2) A suitable non-relativistic limiting procedure 
(for $c^2 \rightarrow \infty$) from the four-dimensional level has then been
utilized. 
Precisely, the limiting procedure has been applied to the 
Einstein-Hilbert-De Witt 
action for the gravitational field plus
a matter action corresponding to a single mass point, under the assumption 
of the existence of a global 3+1 splitting of the total Action, 
and of a suitable
parametrization of the 4-metric tensor in terms of powers of $c^2$. Once the
expansion in powers of $1/c^2$ has been explicitly calculated, we 
have made the {\it
Ansatz} of identifying the basic {\it Galilean Action} {\sl A} 
with the $zero^{th}$ 
order term of the expansion itself. 
3) The resulting Action turned out to be
dependent on 27 fields beside the degrees of freedom of the
mass-point. Eleven           
among these fields are {\it gauge} fields of the Galilei group, having
definite {\it inertial-gravitational} properties, while the remaining 16
fields are not coupled to matter and 
play the role of {\it auxiliary} fields that guarantee 
the general covariance of the theory.

          In force of these results, unlike the case of the four-dimensional 
framework
which does not lend itself to any easy generalization,
a way out of the above constraint appears naturally in our formulation
and is suggested by the very
structure of the total Galilean Action {\sl A}. Indeed, it is natural to
try to eliminate all the {\it auxiliary} fields that are not coupled
to matter. This result is obtained through the new {\it Ansatz} constituted
by putting equal to zero, by hand, as tensor equations,
all the {\it  auxiliary} fields. As a matter of fact, in this way 
we {\it define} a {\it new variational problem} which
turns out to provide consistently the following results: 
(a) The theory contains
only fields coupled to matter. (b) The theory is still {\it gauge-invariant}
(properly {\it quasi-invariant}) under the {\it local} Galilei group.
(c) The condition of Newtonian flatness
($R_{\mu\nu} = 4\pi G \rho(z) t_\mu t_\nu$) no longer appears and
Riemannian three-spaces with non-zero curvature are allowed. 

         It turns out that, within the theory obtained in this way, the 
eleventh gauge field $\Theta(t)$, originally 
generated by the {\it central extension} of the
Galilei group, has no dynamical meaning and can be reabsorbed in the
definition of the {\it absolute time}, so that the theory has {\it ten} 
effective fields. Then, the constraint analysis shows 
that the three-metric $ g_{i j} $ possesses 
three dynamical degrees of freedom. 
The structure of the constraints chains of this theory is rather involved,
and its analysis has not been carried through completely in the present
paper.

A more interesting theory, with {\it eleven} field, is obtained by allowing the
$\Theta$ field to depend on the space variables ${z}$, besides the time ${t}$.
This field, which appears to be a classical analogue of the {\it dilaton} 
field,
gives rise to a quite different constraint structure, which is more like
that of general relativity. It turns out, in fact, that the three-metric 
$ g_{ij} $  has
now {\it two} dynamical degrees of freedom, while all the other fields are
constrained either by gauge conditions corresponding to first-class 
constraints
(as the inertial force vector $ A_k $) or by second-class pairs of constraints
(as the scalar gravitational potential and the {\it dilaton} field). 
From this point of
view, it can be said that this theory shares, so to speak, an intermediate
status between pure Newtonian theory in which there is only the gravitational
"force" associated to the scalar potential $\varphi$, and general 
relativity in
which there is no "force" and the whole dynamical description is provided by
the 4-metric. It is interesting to find  that within a linear approximation of  the
{\it eleven-fields} theory, the dynamical degrees of freedom of the spatial
3-metric show a {\it graviton-like} nature. Indeed, they satisfy an hyperbolic
wave equation and propagate with a velocity related to the square root of the
zero$^{th}$-order weak-field approximation of the scalar Newtonian potential. 
In this way the latter plays the additional role of {\it cosmological}
background. This result seems quite remarkable from both a conceptual 
and a historical point of view. It is worth recalling that Einstein, 
in his first attempts towards a relativistic theory of gravitation, 
introduced a variable speed of light playing
the role of the gravitational potential (see, for example Norton 
\cite{Einstein}).

Section 2 is dedicated to the presentation and discussion of 
the generalized Newtonian gravitational theories as {\it gauge}-invariant
theories of the Galilei group: the {\it ten-fields}-theory (Section 3), and
the {\it eleven-field}-theory (Section 4). Many calculations and special
results are reported in three Appendixes.

 
\setcounter{equation}{0}
\section{Generalized Newtonian Gravities: Galilei {\it gauge}-invariant 
         theories for some  {\it Special Newtonian Manifolds} }

In a previous paper \cite{DePietri} we have shown that it
is possible to implement the standard Newtonian gravity as a 
covariant field theory. The fundamental fields of this theory
are a 3-dimensional Euclidean metric $g_{ij}$, an {\it inertial-gravitational}
vector field $A_i$, a scalar
field $A$ which plays the role of a generalized Newton's potential,
and the {\it time-reparametrization} field $\Theta$. 
It was shown, moreover, that these fields are the {\it gauge} 
fields associated
to the reinterpretation of the Galilei group as a {\it localized} 
group and that they
can be exploited to define the {\it Special Newtonian} space-time structure
on which the four dimensional Cartan's reformulations of
Newtonian Gravity \cite{Cartan} is based. 
The price that we had to pay for the above result was
the introduction of 16 ``auxiliary'' fields,
say $\alpha_0$, $\alpha_i$, $\gamma_{ij}$,  $\beta_{ij}$,
that do not have any physical role besides that of allowing 
a generally covariant formulation. 
Moreover, they do not couple to matter (a mass-point),
and correspond to {\it non-propagating degrees of freedom} in the
standard Newtonian theory. Our reformulation has no physical degree of
freedom and, of course, flat metric. 
         
          We want now to search for a possible true generalization
of Newton theory which, essentially, has to allow for a non-flat metric
and a possible dynamical evolution of it in the spirit of general relativity 
(though, of course, in {\it absolute time}).
As we shall see, contrary to what seems to be a widespread opinion, this 
is in fact realizable. Since we have, so to speak, to {\it reduce} the 
set of conditions which force the flatness and absolute nature of Newton's
space, a natural way for this generalization is already
inscribed in the structure of the 27 fields Newtonian theory, because we have 
here the liberty to try to constrain or even eliminate some or all of the
{\it auxiliary} fields without modifying the variety of fields physically
interacting with matter. After all, only the original eleven fields are
directly connected with the {\it gaugeization} of the Galilei group and, 
furthermore, only these latter are correlated to the geometric 
space-time-Newtonian structures. As a matter of fact, we will derive theories
which remain {\it gauge-invariant} theories of the Galilei group in a peculiar
way. 

           We will adopt here the simplest choice: we will put equal to zero,
by hand, as tensor equations, {\it all} the auxiliary
fields $\alpha_0$, $\alpha_i$, $\gamma_{ij}$,  $\beta_{ij}$. It should be 
clear that, in this way, we are not dealing with, say, a subsector of the
old variational problem, but we are in fact constituting an entirely different
variational problem. In fact, as we shall see, we will obtain a variational
principle for the description of the dynamics of some {\it special 
Newtonian Manifolds} \cite{Kunz} ({\it ten-fields} theory). A different 
theory can be
obtained by allowing the field $\Theta$ to be a function of $t$ {\it and}
$\BMz$. This new theory ({\it eleven-field} theory) describes an additional
{\it dilaton-like} degree of freedom.  Both these theories describe
fields coupled  to matter. It is not clear, however, at the present level
of analysis whether the
formal results are completlly consistent from a distributional point of
view (singularity on the world-line like in the relativistic case of
particles plus fields). Let us remark that, in absence of matter, if
we allow $\alpha_0$ to be different from zero, the resulting theory is
a subsector of the ten-fields one. Probably, in order to fit a
formally consistent theory in presence of matter one has to add extra
couplings of $\alpha_0$ to matter which cannot be obtained by a
limiting procedure from general relativity. Yet, even if we have not 
carried out a
complete analysis of all the possibilities, it is most likely that the
only formally consistent theories without extra coupling to matter are
the {\it ten} and {\it eleven-fields} theories just mentioned.

\newpage
If we keep only the fields that explicitly interact with the
mass-point, i.e. if we set 
$\alpha_0=\alpha_i=\gamma_{ij}=\beta_{ij}=0$, in an {\it arbitrary}
reference frame, the $c^{-2}$ expansion of the total action  
of \cite{DePietri} can be rewritten as: \nopagebreak[4]
\Be
\Ba{rcl}
\tilde{\Sc}   &=&\Sc_F + \Sc_M =  \\[2 mm]
     &=&\Ds \frac{c^3}{16\pi G} \int dtd^3z \sqrt{g} N
                \left[ { {R}
                        + \frac{1}{N^2}
                          g^{ik} g^{jl} (B_{ij}B_{kl} - B_{ik}B_{jl})
                       } \right]  \\[3 mm]
     & &\Ds  - m c \int d\lambda
                 \sqrt{ - g_{\mu\nu} \primato{x}^\mu \primato{x}^\nu } 
\\[4 mm]
     &=&\Ds ~c^4 \left[ {
            \frac{1}{16\pi G} \int dtd^3z \sqrt{g} \Theta  R
                     }\right] 
\\[4 mm]
     & &\Ds +c^2 \left[ {
            \frac{1}{16\pi G} \int dtd^3z \sqrt{g}
                \left[ { - \frac{A}{\Theta} {R}
                 +\Theta g^{ik} g^{jl} (B_{ij}B_{kl} - B_{ik}B_{jl})
                 } \right]
             - m  \int d\lambda \Theta \primato{t}
                  }\right] 
\\[4 mm]
     & &\Ds +~~ \left[ {
            \frac{1}{16\pi G} \int dtd^3z \sqrt{g}
                \left[ { - \frac{A^2}{2\Theta^3} {R}
                        + \frac{A}{\Theta^3}
                          g^{ik} g^{jl} (B_{ij}B_{kl} - B_{ik}B_{jl})
                       } \right] }\right.   \\[3 mm]
     & &\Ds ~\left.{ +m  \int d\lambda {m \over \Theta \primato{t}}
                       \left[ \frac{1}{2}g_{ij}
                              (\primato{x}^i + g^{ik} A_k \primato{t} )
                              (\primato{x}^j + g^{jl} A_l \primato{t} )
                             +A \primato{t} \primato{t}
                       \right]
                   }\right] 
\\[4 mm]
     & &+ O(1/c^2)  ~~~.\\
\Ea
\nome{develop}
\Ee
Now, the zero$^{\rm th}$ order term can be written:
\Be
\Ba{rcl}
 \tilde{\Sc} &=& \tilde{\Sc}_F + \tilde{\Sc}_M \\[2 mm]
         &=&\Ds \frac{1}{16\pi G} \int dtd^3\! z \sqrt{g}
                \left[ { - \frac{A^2}{2\Theta^3} {R}
                        + \frac{A}{\Theta^3}
                          g^{ik} g^{jl} (B_{ij}B_{kl} - B_{ik}B_{jl})
                       } \right]  \\[3 mm]
          & &\Ds + m  \int dtd^3\! z  {m \over \Theta }
                       \left[ \frac{1}{2}g_{ij}
                              (\dot{x}^i + g^{ik} A_k )
                              (\dot{x}^j + g^{jl} A_l )
                             + A  \right]
                       \delta [{\BMz} - {\BMx}(t)] ~~. \\
\Ea
\nome{ActionZero}
\Ee
As in the previous paper \cite{DePietri} we will make the {\it Ansatz} that
the {\it total action} for the generalized Galilean field theory with a
mass-point be the zero$^{th}$ order expression (\ref{ActionZero}). 
The meaning of the symbols here is the following: $g_{ij}$
is a {\it three-dimensional} metric (with signature 3), $A_i$ is an
{\it inertial-gravitational} vector field, $A$ is a
{\it gravitational} scalar field,
and $\Theta$ is the {\it time-reparametrization} field. 
$R$ denotes the three-dimesional scalar
curvature associated to the unique symmetric covariant
derivative $\nabla_i$ compatible with $g_{ij}$, and
$B_{ij}$ is given by:
\Be
 B_{ij} = \frac{1}{2} [\nabla_i A_{j}+\nabla_j A_i 
                          -\partder{g_{ij}}{t} ]
\Ee

The Euler-Lagrange equations
for the mass-point and the fields result:
\Be
\left\{ {
\Ba{rcl}
\EL_{A} &\equiv&\Ds \frac{1}{16\pi G} \frac{\sqrt{g}}{\Theta^3}
                \left[ { - {A} {R}
                        + g^{ik} g^{jl} (B_{ij}B_{kl} - B_{ik}B_{jl})
                       } \right]
             + \frac{m}{\Theta}  \delta^3[{\BMz} - {\BMx}(t)] \deq 0 
\\[3 mm]
\EL_{A_i} &\equiv&\Ds \frac{1}{8\pi G} \frac{1}{\Theta^3}
              \left\{  {
               \partial_j \left[ { - \sqrt{g} {A}
                         [ g^{ik} g^{jl} - g^{ij} g^{kl} ] B_{kl}
                         } \right]   }\right.  \\[3 mm]
           & &\Ds ~ \left.{
                +\left[ { - \sqrt{g} {A}
                         [ g^{rk} g^{sl} - g^{rs} g^{kl} ] B_{kl}
                         } \right] \Gamma^i_{rs}
               } \right\}  \\[3 mm]
           & &\Ds  + \frac{m}{\Theta}  \delta^3 [{\BMz} - {\BMx}(t)]
               ~\left[ \dot{x}^i + g^{ij} A_j \right] \deq 0 \\[3 mm]
 \EL_{\Theta} &\equiv&\Ds  \int d^3\! z \left\{
                \frac{3}{16\pi G}\sqrt{g}
                \left[ {  \frac{A^2}{2\Theta^4} {R}
                        - \frac{A}{\Theta^4}
                          g^{ik} g^{jl} (B_{ij}B_{kl} - B_{ik}B_{jl})
                       } \right]  \right. \\[3 mm]
          & &\Ds \left. ~~~~ - {m \over \theta^2 }
                       \left[ \frac{1}{2}g_{ij}
                              (\dot{x}^i + g^{ik} A_k )
                              (\dot{x}^j + g^{jl} A_l )
                             \right]
                         \delta^3 [{\BMz} - {\BMx}(t)] \right\} 
 \deq 0 \\[3 mm]
 \EL_{g_{ij}} &\equiv&\Ds   {1 \over 16 \pi G} \left\{
                  \sqrt{g} (g^{ir}g^{js} - g^{ij}g^{rs})
                  \nabla_r\nabla_s \left[ {A^2 \over \Theta^3} \right]
                  + {\sqrt{g} A^2 \over 2\Theta^3}
                  [ R^{ij} - \half g^{ij} R]        \right. \\[3 mm]
              & &\Ds  ~~~~+ {2 \sqrt{g} A \over \Theta^3}
                  [B^{ir} B^{js} g_{rs} - B^{ij} {\rm Tr} B]  \\[3 mm]
              & &\Ds  \left.
                  ~~~~+ {d\over dt}
                        \left[ {\sqrt{g} A \over \Theta^3}
                               ( g^{ir} g^{js} - g^{ij} g^{rs} ) B_{rs}
                        \right] \right\} \\[3 mm]
              & &\Ds  + {m \over 2\Theta}
                              (\dot{x}^i + g^{ik} A_k )
                              (\dot{x}^j + g^{jl} A_l )
                      {\delta^3\!}[{\BMz}-{\BMx}(t)] \deq 0  
\\[3 mm]
 \EL_{x^i} &=&  \ddot{x}^i + \Gamma^i_{kl} \dot{x}^k \dot{x}^l  \\[1 mm]
           & & \Ds + { \dot{\Theta} \over \Theta } 
                      \left[ { \dot{x}^i + g^{ij} A_j }\right]
            + g^{ij} \partder{ g_{jl} }{t} \dot{x}^l   \\[1 mm]
           &~& \Ds -  g^{ij} 
            \left[ { \partder{A_0}{x^j} - \partder{A_j}{t} } \right]
             -  g^{ij} 
            \left[ { \partder{A_l}{x^j} - \partder{A_j}{x^l} } \right] 
             \dot{x}^l \deq 0 ~~.
\Ea
}\right.
\nome{EulerLagZero}
\Ee

Let us see that, in this way, as in the case of the standardNewtonian 
theory developed in the previous paper \cite{DePietri},
we have constructed a theory which is invariant under the {\it local}
Galilei group. In that paper it was shown that the localized
Galilei group operations for the mass-point coordinates and for the fields are
naturally defined by the following infinitesimal transformations: 
\Be
\left\{
\Ba{lcl}
\delta ~t(\lambda )  
&=& -\varepsilon (t(\lambda ))   \\
\delta x^i(\lambda )
&=& \varepsilon^i({\BMx},t) 
   - c_{jl}^{~~i} \omega^j ({\BMx},t) x^k 
   - t~v^i({\BMx},t) \\
&=& \tilde{\eta}^i ({\BMx},t)   ~,
\Ea
\right.
\nome{transfparGauA}
\Ee
and
\Be
\left\{
\Ba{rl}
{\delta}~\Theta  &= \dot{\epsilon}(t) \Theta (t)  
\\[2 mm]
{\delta}  g_{ij} &= \Ds 
                  - \partder{\tilde\eta^k ({\BMx},t) }{x^i} g_{kj}
                  - \partder{\tilde\eta^k ({\BMx},t) }{x^i} g_{kj}
\\[2 mm]
{\delta}  A_0     &= \Ds 2 \dot{\varepsilon} A_0
                  - A_i \partder{\tilde\eta^i}{t}
                  - \Theta \partder{}{t}
                  \left[{g_{ij} v^i x^j}\right] \\[2 mm]
{\delta} A_i     &= \Ds \dot{\varepsilon} A_i
                  - A_j \partder{\tilde\eta^j}{x^i}
                  - g_{ij}  \partder{\tilde\eta^j}{t}
                  - \Theta \partder{}{x^i}
                  \left[{g_{ij} v^i x^j}\right]  ~~.
\Ea
\right.
\nome{tranCampGau}
\Ee
where $c_{jk}^{~~i} = \epsilon_{ijk}$ are the usual structure constant
of the O(3) rotation group, $\varepsilon$ is the parameter of
the infinitesimal time-translation, $\varepsilon^i$ are the parameters of
the infinitesimal space-translations, $\omega^i$ are the parameters
of the infinitesimal space rotations, and the $v^i$ are those of the infinitesimal
Galilei boosts.

In fact, if we now adopt these transformation rules, 
the variation of the total action under the transformations
of the mass-point coordinates and of the {\it gauge} fields
(\ref{transfparGauA},\ref{tranCampGau}) results:
\Be
\Ba{rcl}
{\delta} \tilde{\Sc} &=&\Ds  \int dtd^3\! z 
                         \left\{  \dot\varepsilon \tilde{\Lc}
               +\left[ {
                      \frac{1}{16\pi G} \frac{\sqrt{g}}{\Theta^2}
                      \left(  - {A} {R}  + \Gamma \right)
                     + m \delta^3 [{\BMz} - {\BMx}(t)]
                } \right] 
     \left[ \partder{\Fc}{t} - A_r g^{rs} \partder{\Fc}{z^s} \right] 
                 \right. \\[2 mm]
           & &\Ds ~~ -  \frac{1}{8\pi G}          
          \Bigg[ \partder{}{z^j} \left( { 
                           \sqrt{g} \frac{A}{\Theta^2}
                         [ B^{ij}  - ({\rm Tr} B) g^{ij} ]
                              }\right) 
                 + \sqrt{g} \frac{A}{\Theta^2}
                   [ B^{rs}  - ({\rm Tr} B)g^{rs} ] \Gamma^i_{rs}  
                 \\[2 mm]
           & &\Ds ~~  \left. - { m  \delta^3 [{\BMz} - {\BMx}(t)]
                      ~\left( \dot{x}^i + g^{ij} A_j \right)
                      } \Bigg] \partder{\Fc}{z^i}
               + \frac{1}{8\pi G}  \partder{}{z^i}
                        \left( \frac{\sqrt{g} A}{\Theta^2} 
                               [B^{ij} - ({\rm Tr}B) g^{ij} ]
                               \partder{\Fc}{z^j} 
                             \right) \right\} \\[2 mm]
   &=&\Ds  \int dt d^3\! z \left\{ { \dot\varepsilon \tilde{\Lc} 
                     +  \Theta \EL_{A} 
               \left( \partder{\Fc}{t} - A_r g^{rs} \partder{\Fc}{z^s} 
                    \right) 
              +\Theta \EL_{A_i} \partder{\Fc}{z^i}   }\right. \\[2 mm]
           & &\Ds ~~ \left. {
              +\frac{1}{8\pi G} \partder{}{z^i}
                        \left( \frac{\sqrt{g} A}{\Theta^2} 
                               [B^{ij} - ({\rm Tr}B) g^{ij} ]
                               \partder{\Fc}{z^j} 
                               \right)  } \right\} ~~.
\Ea
\nome{variazAzTot}
\Ee 
Therefore we have found the important result that the total
action is {\it quasi-invariant} under the transformations
(\ref{transfparGauA},\ref{tranCampGau}) {\it in force} of the equations
of motion. We can thereby conclude that the theory has a {\it local} Galilei
invariance {\it modulo} the equations of motion. Let us remark that
this peculiarity is precisely what it should be expected in the case
of a variational principle corresponding to a {\it singular} Lagrangian.


\setcounter{equation}{0}
\section{A ten-fields theory}

It is easy to show that, as in the first variational problem of the 
preceding paper ({\it
27-fields} theory), the field $\Theta( t)$ has no real dynamical
content 
also in the variational problem corresponding to the action
(\ref{ActionZero}), since its effect amounts only to a redefinition of
the evolution parameter t in the expression $T(t)=\int_0^t d\tau
\Theta (\tau )$. Indeed, by redefining the fields $A_0$ and $A_i$ as
in the standard Newtonian case of \cite{DePietri}:
\Be
\left\{ {
\Ba{rl}
\tilde{A}_0  &\equiv \Ds {A_0 \over \Theta^2}
                             ~~;\qquad  \tilde{A} 
              \equiv {A   \over \Theta^2}   \\[3 mm]
\tilde{A}_i  &\equiv \Ds {A_i \over \Theta}   \\[3 mm]
\tilde{B}_{ij} &\equiv \Ds  {{B}_{ij}  \over \Theta}
                       =  \frac{1}{2} [ \nabla_i \tilde{A}_{j}
                             + \nabla_j \tilde{A}_i
                             - \partder{g_{ij}}{T} ] ~~, \\
\Ea
}\right.
\nome{TempoN}
\Ee
the total action (\ref{ActionZero}) can be re-written as:
\Be
\Ba{rl}
\tilde{\Sc}  &=\Ds  \int dT L[T]\\[3 mm]
          &= \Ds \frac{1}{16\pi G} \int dTd^3\! z \sqrt{g}
                \left[ { - \frac{\tilde{A}^2}{2} {R}
                        + \tilde{A}
                          g^{ik} g^{jl} (\tilde{B}_{ij}\tilde{B}_{kl}
                                        -\tilde{B}_{ik}\tilde{B}_{jl})
                       } \right]  \\[3 mm]
          &~~\Ds + m  \int dTd^3\! z
                       \left[ \frac{1}{2}g_{ij}
                              ( \frac{d{x}^i}{dT} + g^{ik} \tilde{A}_k )
                              ( \frac{d{x}^j}{dT} + g^{jl} \tilde{A}_l )
                             + \tilde{A}  \right]
                      \delta^3 [{\BMz} - {\BMx}(T)] ~~, 
\Ea
\nome{ActionZeroT}
\Ee
i.e., in a form independent of $\Theta (t)$. Correspondingly, the
Euler-Lagrange equations  are just 
the Eq.(\ref{EulerLagZero}) without $\Theta$.

We shall deal now with the constraint analysis within
the Hamiltonian formalism.
The canonical momenta [$\dot{f}=\partder{f}{T}$] are defined by:
\Be
\left\{
\Ba{rl}
  p_k    &\Ds \equiv \partder{L[T]}{\dot{x^k}}
           = m [ g_{ki} \dot{x}^i + \tilde{A}_k ] \\[3.5 mm]
 \pi^{i} &\Ds \equiv {\delta L \over \delta \dot{\tilde{A}}_{i} } = 0 
\\[3.5 mm]
 \pi_{A} &\Ds \equiv {\delta L[T] \over \delta \dot{\tilde{A}} } = 0 
\\[3.5 mm]
 \pi^{rs}&\Ds \equiv {\delta L[T] \over \delta \dot{g}_{ij} }
          = - {\sqrt{g}\tilde{A} \over 16 \pi G} 
            \left( g^{rk}g^{sl} - g^{rs}g^{kl} \right) \tilde{B}_{kl} 
~~. 
\Ea
\right.
\nome{DefMomenti}
\Ee
Therefore, since the Lagrangian is independent of the {\it velocities}
$\dot{\tilde{A}}$ and $\dot{\tilde{A}}_i $, we have, first of all, the
{\it primary} constraints
\Be
\pi^{i} \simeq 0 ~~~~,\pi_{A} \simeq 0 ~~~.
\nome{vincoliPri}
\Ee
The Dirac Hamiltonian is given by:
\Be
\Ba{rl}
  H_c &= \dot{x}^k p_k
            + \int d^3z \left[
                 {\pi^i \dot{\tilde{A}}_i + \pi_{A} \dot{\tilde{A}}
                  + \pi^{ij} \dot{g}_{ij} }\right]
            - L[T] \\
  H_d &= \int d^3z \left[{ { \tilde{A}^2  \over 32\pi G} \Hc_I
                              +{ 16 \pi G \over \tilde{A}} \Hc_E
                              + \left[ \frac{1}{2m}g^{ij} p_i p_j
                                      - m \tilde{A} \right]
                                        \delta^3 [{\BMz} - {\BMx}(T)]
                             }\right] \\
           &~+\int d^3z \left[{  \tilde{A}_i g^{ij} \phi_j
                               +\pi^i \lambda_i+\pi_{A}{\lambda^{A}}
                             }\right] ~~, \\
\Ea
\nome{HamiltonianaC}
\Ee
where for further convenience we have introduced the following 
notation\footnote{Note that, while the expressions of $\phi$,
             $\Hc_I$ and $\Hc_E$ are identical to the corresponding
             quantities introduced in the usual Hamiltonian
             formalism of general relativity, the ADM 
             super-Hamiltonian is instead $\Hc_\bot = \Hc_I - \Hc_E$.}:
\Be
\left\{
\Ba{rl}
\Hc_I   &= \sqrt{g} R \\[2 mm]
\Hc_E   &= {1 \over \sqrt{g} } 
           [ g_{ik} g_{jl} - \frac{1}{2} g_{ij} g_{kl} ]
           \pi^{ij} \pi^{kl} \\[2 mm]
\phi_i &=   2 g_{ij} \nabla_k \pi^{jk}
           + p_i \delta^3 [{\BMz} - {\BMx}(T)]  ~~. \\
\Ea
\right.
\nome{DefinizioniH}
\Ee
For future use, we list here the relevant algebraic relations
involving the above quantities:
\Be
\Ba{rl}
 \{ \phi_i ({\BMz},T) , \phi_i (\primato{{\BMz}},T) \}
           &=\Ds - \left[ \phi_j ({\BMz},T) \partder{}{z^i}
                     +\phi_i (\primato{\BMz},T) \partder{}{z^j}
              \right] \delta^3 [{\BMz}-\primato{\BMz}] \\[2 mm]
 \{ \phi_i ({\BMz},T) , \Hc_I (\primato{{\BMz}},T) \}
           &=\Ds - \Hc_I(\primato{{\BMz}},T)\partder{}{z^i}
             \delta^3 [{\BMz}-\primato{\BMz}] \\[2 mm]
 \{ \phi_i ({\BMz},T) , \Hc_E (\primato{{\BMz}},T) \}
           &=\Ds - \Hc_E(\primato{{\BMz}},T)
             \partder{}{z^i}\delta^3 [{\BMz}-\primato{\BMz}] \\[2 mm]
 \{ \Hc_I ({\BMz},T) , \Hc_E (\primato{{\BMz}},T) \}
           &=\Ds -2\pi^{rs} \left[ R_{rs} - \frac{1}{4} g_{rs} R \right]
               \delta^3 [{\BMz}-\primato{\BMz}]   \\[2 mm]
            &\Ds ~~+2\pi^{rs} (\primato{{\BMz}},T)
                 \left[{ \partder{^2}{{\BMz}^r \partial {\BMz}_s}
               + \Gamma^k_{rs} (\primato{{\BMz}},T) \partder{}{{\BMz}^k}
                 }\right]  \delta^3 [{\BMz}-\primato{\BMz}] ~, \\
\Ea
\nome{ProprietaH}
\Ee
where $\nabla$ is the covariant derivation with respect to
the metric $g_{ij}$.

Notice that the $\phi_i$'s are the canonical generators
of the coordinate transformations (diffeomorphism group) of
the {\it three-space} with fixed {\it absolute time} since we have
\Be
\Ba{rl}
 \Ds \int d^3\!\primato{z}
 \{ g_{ij} ({\BMz},T) , \phi_k (\primato{{\BMz}},T) \}
 \xi^k(\primato{{\BMz}},T)
     &=\Ds - g_{ij,k} \xi^k - g_{ik} \partder{\xi^k}{z^j}
                         - g_{kj} \partder{\xi^k}{z^i}  \\[3 mm]
 \Ds \int d^3\!\primato{z}
 \{ \pi^{rs} ({\BMz},T) , \phi_k (\primato{{\BMz}},T) \}
 \xi^k(\primato{{\BMz}},T)
     &=\Ds  - \pi^{rs}_{,k} \xi^k - \pi^{rs} \partder{\xi^k}{z^k}
                              + \pi^{rk} \partder{\xi^s}{z^k}
                              + \pi^{ks} \partder{\xi^r}{z^k}  ~~, \\
\Ea
\Ee
a fact which is well-known from the canonical 3+1 formulation
of general relativity \cite{DeWitt67}.

We apply now the Dirac-Bergman procedure. By imposing 
time-conservation of the {\it primary} constraints, we get:
\Be
\Ba{rl}
 \phi_i ({\BMz},T) &=
  - g_{ij}({\BMz},T)  \dot{\pi}^j ({\BMz},T) 
  \equiv - g_{ij}({\BMz},T)   \{ {\pi}^i ({\BMz},T) , H_d \} 
  \simeq 0 \\[2 mm]
 \chi_{A} ({\BMz},T) &\equiv
 \dot{\pi}_{A} ({\BMz},T) = \{ {\pi}_{A} ({\BMz},T) , H_d \} \\[2 mm]
                          &= \Ds {16 \pi G \over \tilde{A}^2} \Hc_E
                            -{\tilde{A} \over 16\pi G} \Hc_I
                            + m   \delta^3\! [{\BMz} - {\BMx}(T)]
                            \simeq 0 ~~,
\Ea
\nome{VincoliSec}
\Ee
which must be imposed as {\it secondary} constraints. 
In turn, time-conservation of these latter gives:
\Be
\Ba{rl}
 \psi_i ({\BMz},T) &\equiv
 \dot{\phi}_i ({\BMz},T) = \{ {\phi}_i ({\BMz},T) , H_d \}  
  \simeq 0  \\[2 mm]
 \psi_{A} ({\BMz},T) &\equiv
 \dot{\chi}_{A} ({\BMz},T) = \{ {\chi}_{A} ({\BMz},T) , H_d \} \\[2 mm]
       &=\Ds  3 \pi^{rs} \left[ R_{rs} - \frac{1}{4} g_{rs} R \right]
         -6 \pi^{rs} {1\over \tilde{A}^2 } \partial_r \tilde{A}
                                           \partial_s \tilde{A} 
\\[2 mm]
       &\Ds ~~+ {16\pi G \over m \tilde{A}^2 \sqrt{g} } \pi^{rs}
          \left[ p_r p_s - \frac{1}{2} g_{rs} g^{lm} p_l p_m  \right]
          \delta^3 [{\BMz} - {\BMx}(T)] \\[2 mm]
       &\Ds ~~
           - {2 \lambda_{A} }
           \left[  {1 \over 32 \pi G} \Hc_I
                 + {16 \pi G \over \tilde{A}^3} \Hc_E \right] 
\simeq 0~. 
\\
\Ea
\nome{VincoliTer}
\Ee
While we have $\psi^i \simeq 0$ as a consequence of the {\it primary}
and {\it secondary}
constraints, the $\psi_{A}$ must be put equal to zero as a further
condition, whose nature must 
be discussed in detail. 

This condition becomes an equation for the multiplier $\lambda_A$,
and the constraints chain stops consequently, if the quantity
\Be
\bar{\chi} \equiv  { 1 \over 32\pi G} \Hc_I
                  +{16 \pi G \over \tilde{A}^3} \Hc_E 
\nome{VincoloMol}
\Ee
is not identically zero. For simplicity, we will discuss only two
particular cases. Actually, a complete treatment of the problem of the
degrees of freedom of the theory would involve a thorough analysis of
the various possible independent constraint sectors corresponding to
different classes of initial conditions \cite{Lusanna}\rlap .

\noindent 
{\bf a)}
The simplest constraint structure obtains for a sector in which 
$\bar\chi {\not\equiv} 0$ everywhere. It is clear that not all the
allowed initial conditions are compatible with this restriction. In
this case we have:

\noindent
{\bf 1)} 3 chains of {\it first class constraints} $\pi^i \simeq 0$, 
$\phi_i \simeq 0$, so that $\tilde{A}_i$ and 3 components of $g_{ij}$
are {\it gauge} variables to be fixed with some {\it gauge fixing},
for instance $\tilde{A}_i\equiv 0$ (Kucha\v{r}'s \cite{Kuch}
{\it non-rotating observer}); \newline {\bf 2)} 1 chain containing a pair of
second class constraints $\pi_A \simeq 0$, $\chi_A \simeq 0$, so that
$\tilde{A}$ is determined by $\chi_{A}\simeq0$ (see
eqs.\ref{VincoliSec}): note that this equation for the potential
hidden inside $\tilde{A}$ is {\it not} Poisson-like. Actually, as we know, it
{\it cannot} be so, in the present conditions since the three-space is
non-flat in general. Its form is precisely the following:
\Be
    2\sqrt{g} \tilde{A} \Delta \tilde{A}
   +2 \sqrt{g} g^{ij}~ \nabla_i\tilde{A} \nabla_j\tilde{A}
           = + 4 \pi m G ~\tilde{A}~\delta^3[{\bf z} - {\bf x}(T)]~.
\nome{Poisson-like}
\Ee
It is interesting to note that, unlike the standard Poisson equation,
Eq.(\ref{Poisson-like}) allows for the solution 
$A \equiv 0$, corresponding to the strong
equation: $\pi^{rs} =0$. This fact could have some interest in connection
with the needs of a non-relativistic cosmology (see for example Rindler
and Friedrichs \cite{RINDE}). 

In this sector there
are 3 physical degrees of freedom in $g_{ij}$ (one more than in the general
relativistic case). The structure of the constraints is illustrated in
Fig.1.\newline
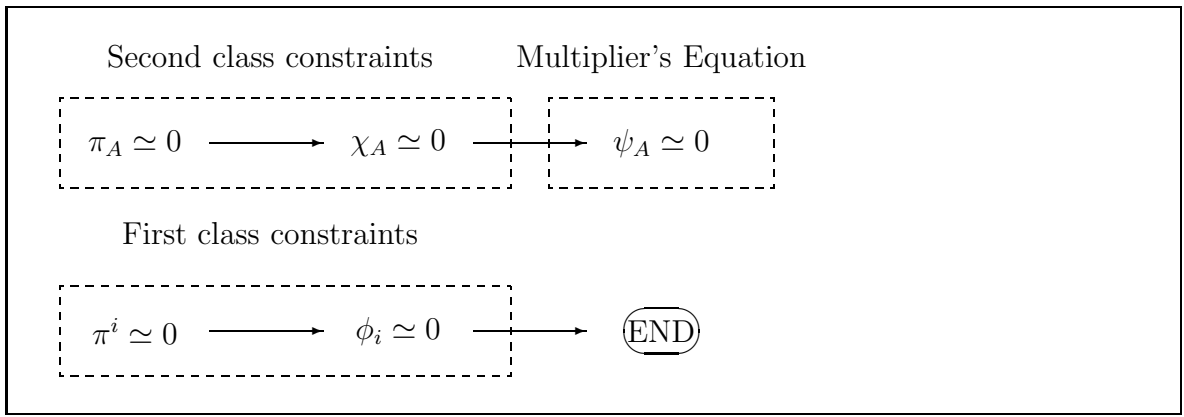
\begin{figure}
\noindent 
\unitlength=1.00mm
\linethickness{0.4pt}
\begin{picture}(1.00,58.00)
\put(20.00,40.00){\makebox(0,0)[cc]{$\pi_A \simeq 0$}}
\put(55.00,40.00){\makebox(0,0)[cc]{$\chi_A \simeq 0$}}
\put(90.00,40.00){\makebox(0,0)[cc]{$\psi_A \simeq 0$}}
\put(30.00,40.00){\vector(1,0){15.00}}
\put(65.00,40.00){\vector(1,0){15.00}}
\put(20.00,15.00){\makebox(0,0)[cc]{$\pi^i \simeq 0$}}
\put(55.00,15.00){\makebox(0,0)[cc]{$\phi_i \simeq 0$}}
\put(90.00,15.00){\makebox(0,0)[cc]{END}}
\put(30.00,15.00){\vector(1,0){15.00}}
\put(65.00,15.00){\vector(1,0){15.00}}
\put(10.00,9.00){\dashbox{1.00}(60.00,12.00)[cb]{}}
\put(10.00,34.00){\dashbox{1.00}(60.00,12.00)[ct]{}}
\put(75.00,34.00){\dashbox{1.00}(30.00,12.00)[ct]{}}
\put(90.00,15.00){\oval(10.00,6.00)[]}
\put(38.00,53.00){\makebox(0,0)[ct]{Second class constraints}}
\put(90.00,53.00){\makebox(0,0)[ct]{Multiplier's Equation}}
\put(38.00,28.00){\makebox(0,0)[cc]{First class constraints}}
\put(3.00,4.00){\framebox(156.00,54.00)[cc]{}}
\end{picture}
\hfill\hfill\break 
\caption{ Constraints chains for the ten-fields}
\end{figure}

\noindent 
{\bf b)}
Let us remark that in the sector $\bar{\chi} \simeq 0$, one has
proliferation \cite{Lusanna} of constraints, i.e. $\psi_A \simeq 0$ is
replaced by $\bar{\chi}\simeq 0$ and $\psi^\prime_A = \psi_A
|_{\chi=0} \simeq 0$. It is extremely difficult to analyze it and we are not
sure that it is self-consistent. Anyway, where it consistent in absence of
matter, it would imply the vanishing of both $\Hc_I$ and $\Hc_E$, as
a consequence of $\chi_A \simeq 0$, $\bar{\chi} \simeq 0$: recall that
the vanishing of both $\Hc_I$ and $\Hc_E$ is a particular solution of
the ADM superHamiltonian constraints also in the general relativistic
case (see footnote 2). Furthermore one should discuss here too the problem
of the central charge $ c^2 \Mc + c^4 \Nc$, mentioned at the end of 
Section 6 of the previous work.

\setcounter{equation}{0}
\subsection{An eleven-fields theory}

Let us now generalize the field $\Theta(t)$  to an expression
depending on time {\it and} on the space coordinates: $\Theta =
\Theta(\BMz,t)$. This generalization is not quite natural from a
Galilean point of view since it introduces {\it dilaton-like} 
\footnote{Allowing this generalization is tantamount to perform a
{\it conformal} transformation on the original four-dimensional metric.
In this way, the opening-out of the light-cones taking place through
the limiting procedure
no longer occur {\it uniformly} across the four-dimensional manifold.}
degrees
of freedom into a Newtonian framework. In this case, of course, 
the field $\Theta$ cannot be
reabsorbed and the total action (\ref{ActionZero}) can be written
as: 
\Be
\Ba{rl}
\tilde{\Sc}  &= \Ds \int dt L[t] \\[2 mm]
          &=\Ds  \frac{1}{16\pi G} \int dtd^3\! z \sqrt{g}
                \left[ { - \frac{{A}^2}{2\Theta^3} {R}
                 + \frac{A}{\Theta^3}
                 g^{ik} g^{jl} ({B}_{ij}{B}_{kl}-{B}_{ik}{B}_{jl})
                 } \right]  \\[2 mm]
          &\Ds ~~+ \int dtd^3\! z \frac{m}{\Theta}
                       \left[ \frac{1}{2}g_{ij}
                              (\dot{x}^i + g^{ik} A_k )
                              (\dot{x}^j + g^{jl} A_l )
                             + A
                             \right]
                       \delta^3 [{\BMz} - {\BMx}(t)] ~~. \\
\Ea
\nome{ActionZeroU}
\Ee

Here again, we shall deal with the problem of investigating the true
dynamical degrees of freedom of the theory by means of a constraint
analysis within the Hamiltonian formalism. In this case too is
profitable to adopt $\tilde{A}=A / \Theta^2$ as a dynamical variable.
The Action then becomes:
\Be
\Ba{rl}
  \tilde{\Sc} &=\Ds \int dt L[t] \\[3 mm]
        &=\Ds \frac{1}{16\pi G} \int dtd^3\! z \sqrt{g}
                \left[ { - \frac{\tilde{A}^2}{2} \Theta {R}
                  + \frac{\tilde{A}}{\Theta}
                  g^{ik} g^{jl} ({B}_{ij}{B}_{kl}-{B}_{ik}{B}_{jl})
                  } \right]  \\[3 mm]
         &\Ds~~+ \int dtd^3\! z \frac{m}{\Theta}
                       \left[ \frac{1}{2}g_{ij}
                              (\dot{x}^i + g^{ik} {A}_k )
                              (\dot{x}^j + g^{jl} {A}_l )
                             + \Theta^2 \tilde{A}
                             \right]
                 \delta^3 [{\BMz} - {\BMx}(t)] ~~. \\
\Ea
\nome{ActionU}
\Ee
The Euler-Lagrange equation of this Action are reported in
appendix A.

The canonical momenta, [$\dot{f}=\partder{f}{t}$] result:
\Be
\left\{
\Ba{rl}
  p_k    &\equiv \Ds \partder{L}{\dot{x^k}}
           = {m \over \Theta} [ g_{ki} \dot{x}^i + {A}_k ] \\[3 mm]
 \pi^{i} &\equiv \Ds {\delta L[t] \over \delta \dot{A}_{i} } = 0 
\\[3 mm]
 \pi_{\Theta} &\equiv \Ds {\delta L[t] \over \delta \dot{\Theta} } = 0
\\[3 mm]
 \pi_{A} &\equiv \Ds {\delta L[t] \over \delta \dot{\tilde{A}} } = 0 
\\[3 mm]
 \pi^{rs}&\equiv \Ds {\delta L[t] \over \delta \dot{g}_{ij} }
          = - {\sqrt{g} \tilde{A} \over 16 \pi G \Theta } 
            \left( g^{rk}g^{sl} - g^{rs}g^{kl} \right) {B}_{kl} ~~. 
\Ea
\right.
\nome{DefMomentiU}
\Ee
Therefore, since the Lagrangian is independent of the {\it
velocities} $\dot{{A}}_i$, $\dot{\Theta}$ and $\dot{\tilde{A}}$, we
have first of all the {\it primary} constraints
\Be
\left\{
\Ba{rcl}
\pi^{i}      &\simeq& 0       \\
\pi_{\Theta} &\simeq& 0       \\
\pi_{A}      &\simeq& 0 ~~~.  \\
\Ea
\right.
\nome{vincoliPriU}
\Ee
The Dirac Hamiltonian is given by:
\Be
\Ba{rl}
  H_c &=\Ds \dot{x}^k p_k
            + \int d^3z \left[
                 {\pi^i \dot{{\tilde{A}}}_i + \pi_{A} \dot{\tilde{A}}
                  + \pi^{ij} \dot{g}_{ij} }\right]
            - L[t] \\[3 mm]
  H_d &=\Ds \int d^3z \Theta
                        \left[{ { \tilde{A}^2  \over 32 \pi G} \Hc_I
                              +{ 16 \pi G \over \tilde{A}} \Hc_E
                              +\left[ \frac{1}{2m}g^{ij} p_i p_j
                                      - m\tilde{A} \right]
                                        \delta^3 [{\BMz} - {\BMx}(t)]
                             }\right] \\[3 mm]
          &\Ds ~+\int d^3z \left[{ - {A}_i g^{ij} \phi_j
                               +\pi^i      \lambda_i
                               +\pi_{A}    \lambda^A
                               +\pi_\Theta \lambda^{\Theta}
                             }\right] ~~, \\
\Ea
\nome{HamiltonianaCU}
\Ee
where notation and algebraic properties of the quantities involved
are the same of eqs.(\ref{DefinizioniH}) and (\ref{ProprietaH}).

We apply now the Dirac-Bergmann procedure. By imposing
time-conservation of the primary constraints, we get:
\Be
\left\{
\Ba{rcl}
 \phi_i ({\BMz},t)  &\equiv &
 - g_{ij} ({\BMz},t) \dot{\pi}^j ({\BMz},t) = 
 - g_{ij} ({\BMz},t) \{ {\pi}^j ({\BMz},t) , H_d \} \\[3 mm]
 &=&\Ds  2 g_{ij} \nabla_k \pi^{jk}
         + p_i \delta^3\! [{\BMz} - {\BMx}(t)]
 \simeq 0 \\[3 mm]
 \chi_{\Theta} ({\BMz},t) & \equiv &
 \dot{\pi}_{\Theta} ({\BMz},t) 
                  = \{ {\pi}_{\Theta} ({\BMz},t) , H_d \} \\[3 mm]
                 &=&\Ds -{ \tilde{A}^2  \over 32\pi G} \Hc_I
                     -{ 16 \pi G \over \tilde{A}} \Hc_E
                     - \left[ \frac{1}{2m}g^{ij} p_i p_j
                     - m \tilde{A} \right]
                     \delta^3\! [{\BMz} - {\BMx}(t)]  \simeq 0 \\[3 mm]
 \chi_{A} ({\BMz},t) &\equiv&
 \dot{\pi}_{A} ({\BMz},t) =\Ds \{ {\pi}_{A} ({\BMz},t) , H_d \} \\[3 mm]
        &=&\Ds \Theta \left\{ {16 \pi G \over \tilde{A}^2} \Hc_E
        -{\tilde{A} \over 16\pi G} \Hc_I
        + m  \delta^3\! [{\BMz} - {\BMx}(t)] \right\}  \simeq 0  ~~,\\
\Ea
\right.
\nome{VincoliSecU}
\Ee
which must be imposed as {\it secondary} constraints. In turn,
time-conservation of these latter, implies that the following weak
equations be satisfied on  the constraint hypersurface:
\Be
\left\{
\Ba{rcl}
\dot{\phi}_i ({\BMz},t) &=& \{ {\phi}_i ({\BMz},t) , H_d \} \\[3 mm]
       &=&-\chi_A \tilde{A}_{,i} - \chi_\Theta \tilde{\Theta}_{,i}
          + \phi_l g^{lk} A_{k,i}
          + A_i \partial_k [g^{lk}\phi_k] \simeq 0  
\\[3 mm]
 \dot{\chi}_{\Theta} ({\BMz},t) 
      &=& \{ {\chi}_{\Theta} ({\BMz},t) , H_d \} 
\\[3 mm]
       &\simeq&   6 \pi^{rs} \partial_r \tilde{A} \partial_s \Theta 
           + 3 \pi^{rs} \Theta \nabla_r\nabla_s \tilde{A} 
          +\tilde{A}\Theta \partial_k[\nabla_l{\pi^{kl}}] 
\\[3 mm]
       & &\Ds -2\tilde{A}\Theta \nabla_l\pi^{kl} 
           \left[ \frac{\Theta_{,k}}{\Theta}
                 -\frac{\tilde{A}_{,k}}{\tilde{A}}
           \right]
\\[3 mm]
       & &+\Ds \frac{p_l g^{lk}}{m} \Hc_M \Theta_{,k}
          +\partial_k \left[ \frac{p_l g^{lk}}{m} \Hc_M \Theta
                      \right]
           \simeq 0 ~~, 
\\[3 mm]
 \dot{\chi}_{A} ({\BMz},t)&=& \{ {\chi}_{A} ({\BMz},t) , H_d \} \\[3 mm]
    &\simeq&\Ds 3 \Theta \pi^{rs} 
    \left[ R_{rs} - \frac{1}{4} g_{rs} R \right]
    -6 \pi^{rs} {\Theta^2\over \tilde{A}^2 } \partial_r \tilde{A}
    \partial_s \tilde{A}
    -3\pi^{rs} \Theta \nabla_r\nabla_s \Theta \\[3 mm]
    & &\Ds + { 16\pi G \Theta^2 \over m \sqrt{g} \tilde{A}^2 } \pi^{rs}
    \left[ p_r p_s - \frac{1}{2} g_{rs} g^{lm} p_l p_m  \right]
    \delta^3\! [{\BMz} - {\BMx}(t)] \\[3 mm]
    & &\Ds  + \frac{2}{\Theta\tilde{A}^2} \Hc_M 
    \left[ \lambda_{A} - A_i g^{ij} \tilde{A}_{,j} \right]  
    -2 \Theta^2 \nabla_l\pi^{kl} 
    \left[ \frac{\Theta_{,k}}{\Theta}
    -\frac{\tilde{A}_{,k}}{\tilde{A}}
    \right]
  \simeq 0  ~~, \\[3 mm]
\Ea
\right.
\nome{VincoliTerUA}
\Ee
where we have defined 
$\Hc_M(\BMz) = ( \frac{1}{2m}g^{ij}p_i p_j - m\tilde{A} ) 
\delta^3[\BMz-\BMx(t)]$.

Now, since $\psi_i \equiv \dot{\phi}_i \simeq 0$ already holds as a
consequence of the {\it primary} and {\it secondary} constraints, the
condition that the remaining weak equations be satisfied amounts to
imposing the following {\it tertiary} constraints:
\Be
\left\{
\Ba{rcl} 
 \psi_{\Theta} ({\BMz},t) &\equiv& \dot{\chi}_\Theta \simeq 0  \\
 \psi_{A} ({\BMz},t) &\equiv& \dot{\chi}_A \simeq 0  ~~.
\Ea
\right.
\nome{VincoliTerU}
\Ee
Let us note that the second of eqs.(\ref{VincoliTerUA}) determines the
Dirac multipliers $\lambda_A$ on the particle world-line. Again, by
imposing time-conservation of the constraints (\ref{VincoliTerU}), we
finally obtain the {\it quaternary} constraints  $\xi_A ({\BMz},t)$
and $\xi_\Theta ({\BMz},t)$ whose esplicit
form is reported in Appendix B (Eqs. B.1 and B.2).

Eqs.(\ref{VincoloQuaUa},\ref{VincoloQuaUb}) allow in
principle to solve for the multipliers
$\lambda_{\Theta}$ and $\lambda_{A}$ and to close the
constraints chains. Since
eqs.(\ref{VincoloQuaUa},\ref{VincoloQuaUb}) are partial
differential equations for $\lambda_\Theta$ and
$\lambda_A$ this is a non-trivial problem which could
be possibly connected with the presence of {\it residual} gauge
degrees of freedoms.
Looking at the algebraic relations existing among all
the constraints, we see that the only {\it first-class}
ones are:
\Be
\left\{
\Ba{rl}
\pi^i  &\simeq 0 \\
\tilde{\phi}_i = {\phi}_i -\pi_A \tilde{A}_{,i} 
                  - \pi_\Theta \Theta_{,i} 
                  +\pi^k_{,k} {A}_i 
                  + \pi^k ( {A}_{i,k} - {A}_{k,i} ) 
          &\simeq 0 ~~.\\
\Ea
\right.
\nome{PrimaClasseU}
\Ee

The whole constraints chains are summarized in Fig. 2, while
the complete constraints {\it algebra} is given in Appendix C.
\par
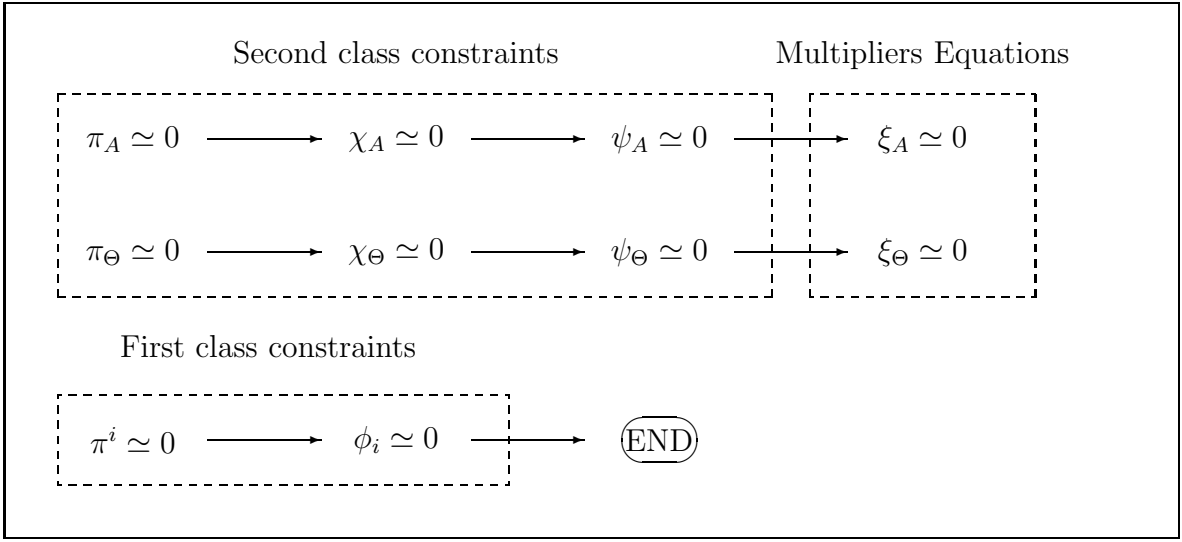
\begin{figure}
\noindent 
\unitlength=1.00mm
\linethickness{0.4pt}
\begin{picture}(1.00,78.00)
\put(20.00,60.00){\makebox(0,0)[cc]{$\pi_A \simeq 0$}}
\put(55.00,60.00){\makebox(0,0)[cc]{$\chi_A \simeq 0$}}
\put(90.00,60.00){\makebox(0,0)[cc]{$\psi_A \simeq 0$}}
\put(125.00,60.00){\makebox(0,0)[cc]{$\xi_A \simeq 0$}}
\put(30.00,60.00){\vector(1,0){15.00}}
\put(65.00,60.00){\vector(1,0){15.00}}
\put(100.00,60.00){\vector(1,0){15.00}}
\put(20.00,45.00){\makebox(0,0)[cc]{$\pi_\Theta \simeq 0$}}
\put(55.00,45.00){\makebox(0,0)[cc]{$\chi_\Theta \simeq 0$}}
\put(90.00,45.00){\makebox(0,0)[cc]{$\psi_\Theta \simeq 0$}}
\put(125.00,45.00){\makebox(0,0)[cc]{$\xi_\Theta \simeq 0$}}
\put(30.00,45.00){\vector(1,0){15.00}}
\put(65.00,45.00){\vector(1,0){15.00}}
\put(100.00,45.00){\vector(1,0){15.00}}
\put(20.00,20.00){\makebox(0,0)[cc]{$\pi^i \simeq 0$}}
\put(55.00,20.00){\makebox(0,0)[cc]{$\phi_i \simeq 0$}}
\put(90.00,20.00){\makebox(0,0)[cc]{END}}
\put(30.00,20.00){\vector(1,0){15.00}}
\put(65.00,20.00){\vector(1,0){15.00}}
\put(10.00,14.00){\dashbox{1.00}(60.00,12.00)[cb]{}}
\put(10.00,39.00){\dashbox{1.00}(95.00,27.00)[ct]{}}
\put(110.00,39.00){\dashbox{1.00}(30.00,27.00)[ct]{}}
\put(90.00,20.00){\oval(10.00,6.00)[]}
\put(55.00,73.00){\makebox(0,0)[ct]{Second class constraints}}
\put(125.00,73.00){\makebox(0,0)[ct]{Multipliers Equations}}
\put(38.00,31.00){\makebox(0,0)[cb]{First class constraints}}
\put(3.00,7.00){\framebox(156.00,71.00)[cc]{}}
\end{picture}
\hfill\hfill\break 
\caption{  Constraints chains for the eleven-fields theory.}
\end{figure}
While the Gauss constraints $\phi_i \simeq 0$ correspond to  the
pure {\it gauge} nature of three degrees of freedom of $g_{ij}$
(that have to be fixed by {\it three coordinate conditions}),  the
constraints $\pi^i \simeq 0$ correspond in turn to the {\it gauge}
nature of the  fields $\tilde{A}_i$.

Although a complete discussion of the role of the {\it second class} 
constraints in restricting the number of degrees of freedom  could
be carried out only in connection to the properties of  the solutions
of Eq.(\ref{VincoloQuaUa},\ref{VincoloQuaUb}) 
for the Dirac multipliers, it is
reasonable to expect six generally {\it second class} constraints, as  shown
in Fig.2: they fix the fields $\Theta(\BMz,t)$, $\tilde{A}(\BMz,t)$
and one of the remaining degrees of freedom of $g_{ij}(\BMz,t)$ as
functionals of all the other fields \cite{LSecClass}. Therefore we
are left with {\it two} degrees of freedom  of the {\it
three-metric}, as in general relativity. It is interesting to see
that these degrees of freedom are "graviton-like", and that their
propagation properties can be explicitly  exhibited by means of a
linear approximation, in a region far from matter. 

In order to show this, we choose a {\it gauge} fixing chain 
of the {\it gauge} variables associated to the 
{\it first-class} constraints in such a way that
the first ones are just the "harmonic coordinate-conditions".
Precisely, following \cite{JAP}, we start adding 
to the  {\it first class} secondary  constraints 
$\phi_i \simeq 0$~, the gauge fixing constraints
\Be
 \Omega_{k} \equiv g_{kl} \Gamma^l_{rs} g^{rs} \simeq 0
~~.
\Ee
Then, the condition 
\Be
 M_k \equiv \dot{\Omega}_k \simeq 0 
~~,
\nome{8.30}
\Ee 
provides the {\it gauge-fixing} for the {\it first class}
primary constraints $\pi^i \simeq 0$. The fields $A_i$ are determined
by these equations. The time-derivatives of the constraints
(\ref{8.30})
\Be 
\dot{M_k}\simeq 0
~~,
\Ee
fix in turn the multipliers $\lambda_i$. Clearly, the local Galilei 
invariance is thereby broken.

The explicit calculations of the linear approximation
will be worked out by starting with a {\it weak-field} 
approximation for the fields, based on the following
{\it Ansatz} for the zero$^{\rm th}$-order
terms:
\Be
\left\{
\Ba{rccll}
  g_{ij}(\BMz,t)     &=& \delta_{ij} 
                       & + \epsilon~ h_{ij}(\BMz,t)
                       & + ~O(\epsilon^2) \\
  \Theta(\BMz,t)     &=& K_\Theta        
                       & +~\epsilon\Theta^{[1]} (\BMz,t)    
                       & + ~O(\epsilon^2) \\
  \tilde{A} (\BMz,t) &=& - K_A           
                       & + \epsilon~ \tilde{A}^{[1]} (\BMz,t) 
                       & + ~O(\epsilon^2) \\
  A_i         (\BMz,t) &=& 0           
                       & + \epsilon~ {A}_i^{[1]} (\BMz,t) 
                       & + ~O(\epsilon^2) ~~,  \\
\Ea
\right.
\nome{8.41}
\Ee
where $K_A$ and $K_\Theta$ are positive real constants.
Moreover, the tensor $h_{ij}$ will be decomposed 
in the usual {\it transverse-traceless} form:
\Be
 h_{ij} = h^{TT}_{ij} 
         +\frac{1}{2}\left( \delta_{ij} h^T - (\Delta^{-1} h^T)_{,ij}
                     \right)
         + h^L_{i,j} + h^L_{j,i}   
~~.
\nome{8.42}
\Ee
Given (\ref{8.41}), we obtain the following week-field expressions
for the non-vanishing canonical momenta 
\Be
\left\{
\Ba{rcl}
 \pi^{ij} &=& \Ds \epsilon~ \frac{1}{32\pi G}\frac{K_A}{K_\Theta} 
           [ \delta^{ir}\delta^{js}-\delta^{ij}\delta^{rs} ]
           [ A_{r,s}^{[1]} + A_{s,r}^{[1]} - \dot{h}_{rs} ]  
           + O(\epsilon^2) 
\\[2 mm]
 \pi_\Theta &=& 0 \\[1 mm] 
 \pi_A      &=& 0 \\[1 mm] 
 \pi_i      &=& 0 ~~.
\Ea
\right.
\nome{8.45}
\Ee
From equations (\ref{8.41}-\ref{8.42}), we get the following expansion for the
constraints:
\begin{eqnarray*}
  \chi_A      &\simeq& \Ds -\frac{K_\Theta K_A}{16 \pi G}
                       \epsilon \nabla h^T + O(\epsilon^2)  \\  
  \chi_\Theta &\simeq& \Ds \frac{K_A^2}{16 \pi G}
                       \epsilon \nabla h^T + O(\epsilon^2)  \\  
  \psi_A      &\simeq& 0 + O(\epsilon^2)  \\  
  \psi_\Theta &\simeq& 0 + O(\epsilon^2)  \\  
  \Omega_k    &\simeq& \Ds \epsilon \left[
                 \frac{1}{2} \partial_k h^T + \nabla h^L_k
                 \right] + O(\epsilon^2)  \\
  \phi_k      &\simeq& \Ds \Ds \epsilon 
               \frac{K_A}{16 \pi G K_\Theta}
               \left[ \partial_k h^T + \partial^r \left(
                 A^{[1]}_{k,r} - A^{[1]}_{r,k}
                -h^{L}_{k,r} + h^{L}_{r,k}
               \right)\right]
              + O(\epsilon^2) \\
  M_k         &\simeq& 
\epsilon M^{[1]}_k [K_A,K_\Theta;A^{[1]}_k,h_k^{L},h^T ] 
                     + O(\epsilon^2) ~~~.  
\end{eqnarray*}
The vanishing of the terms of order $\epsilon$ determines
the quantities $h^T, h^L_k, A^{[1]}_i$ while
leaves the quantities 
$h^{TT}_{ij}, \Theta^{[1]}, \tilde{A}^{[1]}$ undetermined. 

Now, let us wright the
multipliers  $\lambda^\Theta$, $\lambda^A$,
$\lambda_i$ as power series in $\epsilon$:
\Be
\left\{
\Ba{rcl}
  \lambda^\Theta(\BMz,t) &=& \lambda_\Theta^{[0]} (\BMz,t) 
                             + \epsilon \lambda_\Theta^{[1]} (\BMz,t) 
                             + O(\epsilon^2) \\
  \lambda^A(\BMz,t)      &=& \lambda_A^{[0]} (\BMz,t) 
                             + \epsilon \lambda_A^{[1]} (\BMz,t) 
                             + O(\epsilon^2)  ~~,\\
  \lambda_i(\BMz,t)      &=& \lambda_i^{[0]} (\BMz,t) 
                             + \epsilon \lambda_i^{[1]} (\BMz,t) 
                             + O(\epsilon^2)  ~~.\\
\Ea
\right.
\Ee
The equations (\ref{8.30},\ref{VincoloQuaUa},\ref{VincoloQuaUa}) 
for the Dirac multipliers become: 
\begin{eqnarray}
\xi^A     &=& 0 +\epsilon \frac{3 {K_A} }{32\pi G} 
           [ \delta^{ir}\delta^{js}-\delta^{ij}\delta^{rs} ]
           [ A_{r,s}^{[1]} + A_{s,r}^{[1]} - \dot{h}_{rs} ]  
           \partial_r\partial_s \lambda_\Theta^{[0]}
              +O(\epsilon^2) \simeq 0
\nonumber \\[3 mm]
\xi^\Theta&=& 0 + \epsilon \frac{3 {K_A} }{32\pi G} 
           [ \delta^{ir}\delta^{js}-\delta^{ij}\delta^{rs} ]
           [ A_{r,s}^{[1]} + A_{s,r}^{[1]} - \dot{h}_{rs} ]  
           \partial_r\partial_s \lambda_A^{[0]}
              +O(\epsilon^2) \simeq 0
\nome{VINi} \\[3 mm]
\dot{M}_k &=& - \Delta \lambda^{[0]}_{,k} +
              \epsilon~ \left(
               \frac{1}{2} K_\Theta^2 \Delta \tilde{A}^{[1]}_{,k} 
              -\frac{1}{2} K_\Theta K_A  \Delta \Theta^{[1]}_{,k}
              -\frac{1}{8} K_\Theta K_A^2 \Delta  h^T_{,k}
              - \Delta \lambda^{[1]}_{k}
              \right) 
            +O(\epsilon^2) \simeq 0 ~~.
\nonumber 
\end{eqnarray}
From these equations, it is seen that the {\it Ansatz} (\ref{8.41})
is indeed consistent since Eqs.(3.16) admit the solutions 
$\lambda^{[0]}_\Theta (\BMz,t) =0 $,
$\lambda^{[0]}_A (\BMz,t) = 0 $, 
$\lambda^{[0]}_i (\BMz,t) = 0 $ at the $zero^{th}$ order in $\epsilon$.

Then, the Hamilton equations of motion for $\pi^{ij}$ result:
\Be
\Ba{rcl}
\dot\pi^{ij} &=& \{ \pi^{ij} , H_d \} \\
    &=&
\Ds \epsilon~ \frac{
           [ \delta^{ir}\delta^{js}-\delta^{ij}\delta^{rs} ]
           }{16\pi G}
           \bigg[
           -K_\Theta K_A^2 \frac{1}{2}\Delta h^{TT}_{rs}
           +K_\Theta K_A^2 \frac{1}{4}\Delta h^{T}_{,rs} \\
& & \Ds ~~~~~~~~~~~
           -K_A^2\partial_r \partial_s \Theta^{[1]} 
           -2 K_A K_\Theta\partial_r \partial_s \tilde{A}^{[1]}
           \bigg] 
\Ds+~O(\epsilon^2)  ~~.
\Ea
\nome{8.49}
\Ee
On the other hand, from time differentiation of eqs.(\ref{8.45}), one
also gets
\Be
\Ba{rcl}
\dot\pi^{ij} &=& 
\epsilon \Ds 
    \frac{K_A [ \delta^{ir}\delta^{js}-\delta^{ij}\delta^{rs} ] 
         }{16\pi G K_\Theta} 
           \left[ \lambda^{[1]}_{r,s} + \lambda^{[1]}_{s,r} 
            -\frac{d^2}{dt^2} \left[
                 h^{TT}_{ij} 
                 +\frac{1}{2}\left( 
                     \delta_{ij} h^T - \Delta^{-1} h^T_{,ij}
                     \right)
                 + h^L_{i,j} + h^L_{j,i}   
              \right] 
           \right]  ~~.
\Ea
\nome{8.50}
\Ee
Finally, by confronting (\ref{8.49}) and (\ref{8.50})  (collecting
the expression  $\frac{-\epsilon K_A^2K_\Theta}{32\pi G}  [
\delta^{ir}\delta^{js}-\delta^{ij}\delta^{rs} ]$), inserting in
(\ref{8.50}) the expression for $\lambda_i^{[1]}$ which comes out
from the last equation (\ref{VINi}) at the first order, and
separating out the {\it transverse traceless}, the {\it trace} and 
{\it longitudinal} parts of the resulting equations,  respectively,
it follows:
\Be
\left\{
\Ba{rcl}
  \Ds \frac{d^2}{dt^2} h^T 
&=& 0 +~O(\epsilon) 
\\[3 mm]
  \Ds \frac{d^2}{dt^2} h_i^{L} 
&=& 0 +~O(\epsilon) 
\\[3 mm]
 \Ds  \frac{2}{K_\Theta^2 K_A} \frac{d^2}{dt^2} h^{TT}_{ij} 
&=& \Ds \nabla h^{TT}_{ij} 
+~O(\epsilon) 
~~.
\\[1 mm]
\Ea
\right.
\nome{8.last}
\Ee
Note that the first two equations (\ref{8.last}) are compatible
with the constraints for $h^T, h^L_i$.
The third one is the wave-equation for the "graviton-like" 
degrees of freedom of the three-metrics which, consequently, 
propagate with a velocity given by:
\Be
 V  = K_\Theta \sqrt{\frac{K_A}{2}}  = 
    = \Theta^{[0]} \sqrt{\frac{\tilde{A}^{[0]}}{2}}  
    = \sqrt{\frac{-A_0^{[0]}}{2}}
~~.
\Ee
It is seen that, under the above conditions, the potential $A_0$, at
the lowest order, assumes, as it were, the role of a cosmological
background. At this order, instead, $\Theta^{[1]}$ and 
$\tilde{A}^{[1]}$ remain undetermined.
This result seems to us quite remarkable from both a conceptual
and a historical point of view. 
It is worth recalling that Einstein,
in his first attempts towards a relativistic theory
of gravitation, introduced a variable speed of light playing the
role of the gravitational potential (see Ref. \cite{Einstein}).

\section{Acknowledgments}

Roberto De Pietri wishes to thank C. Rovelli, Al Janis and E.T. Newman for
the hospitality kindly offered to him at the Department of Physics 
and Astronomy of the University of Pittsburgh. Massimo Pauri would like
to express his deep appreciation and thanks to the {\it Center for 
Philosophy of Science}, for the warm and stimulating intellectual
atmosphere experienced there, and the generous partial support
obtained during the completion of the present work at the
Pittsburgh University.




\renewcommand{\theequation}{A.\arabic{equation}}\setcounter{equation}{0}
\section*{Appendix A: The Euler-Lagrange equations of the eleven 
fields theory.}

The Euler-Lagrange equations of the eleven fields theory result:
\Be
\left\{ {
\Ba{rcl}
\EL_{A} &=&\Ds \frac{1}{16\pi G} \frac{\sqrt{g}}{\Theta^3}
                \left[ { - {A} {R}
                        + g^{ik} g^{jl} (B_{ij}B_{kl} - B_{ik}B_{jl})
                       } \right]
      + \frac{m}{\Theta}  \delta^3 [{\BMz} - {\BMx}(t)] \deq 0 \\[3 mm]
\EL_{A_i} &=&\Ds\frac{1}{8\pi G}
              \left\{  {
               \partial_j \left[ { - \frac{1}{\Theta^3} \sqrt{g} {A}
                         ( g^{ik} g^{jl} - g^{ij} g^{kl} ) B_{kl}
                         } \right] }\right. \deq 0 \\[3 mm]
          & &\Ds \left. {
               + \left[ { - \frac{1}{\Theta^3} \sqrt{g} {A}
                         ( g^{rk} g^{sl} - g^{rs} g^{kl} ) B_{kl}
                         } \right] \Gamma^i_{rs}
               } \right\}  \\[3 mm]
           & &\Ds + \frac{m}{\Theta}  \delta^3 [{\BMz} - {\BMx}(t)]
               ~\left[ \dot{x}^i + g^{ij} A_j \right] \deq 0 \\[3 mm]
 \EL_{\Theta} &=&\Ds  \frac{3}{16\pi G}\sqrt{g}
                \left[ {  \frac{A^2}{2\Theta^4} {R}
                        - \frac{A}{\Theta^4}
                          g^{ik} g^{jl} (B_{ij}B_{kl} - B_{ik}B_{jl})
                       } \right]   \\[3 mm]
               & &\Ds  ~~~~ -  {m \over \Theta^2 }
                       \left[ \frac{1}{2}g_{ij}
                              (\dot{x}^i + g^{ik} A_k )
                              (\dot{x}^j + g^{jl} A_l )
                             + A
                             \right]
                   \delta^3 [{\BMz} - {\BMx}(t)] \deq 0 ~~. \\[3 mm]
 \EL_{g_{ij}} &=&\Ds   {1 \over 16 \pi G} \left\{
                   \sqrt{g} (g^{ir}g^{js} - g^{ij}g^{rs})
                   \nabla_r\nabla_s \left[ {A^2 \over \Theta^3} \right]
                   + {\sqrt{g} A^2 \over 2\Theta^3}
                   [ R^{ij} - \half g^{ij} R]        \right. \\[3 mm]
              & &\Ds ~~~~+ {2 \sqrt{g} A \over \Theta^3}
                   [B^{ir} B^{js} g_{rs} - B^{ij} {\rm Tr} B]  \\[3 mm]
              & &\Ds \left.
                  ~~~~+ {d\over dt}
                        \left[ {\sqrt{g} A \over \Theta^3}
                               ( g^{ir} g^{js} - g^{ij} g^{rs} ) B_{rs}
                        \right] \right\} \\[3 mm]
              & &\Ds + {m \over 2\Theta}
                              (\dot{x}^i + g^{ik} A_k )
                              (\dot{x}^j + g^{jl} A_l )
                      {\delta^3\!}[{\BMz}-{\BMx}(t)] \deq 0 
\\[3 mm]
 \EL_{x^i} &=& \ddot{x}^i + \Gamma^i_{kl} \dot{x}^k \dot{x}^l \\[1 mm]
           & & \Ds + { \dot{\Theta} \over \Theta } 
                      \left[ { \dot{x}^i + g^{ij} A_j }\right]
              + g^{ij} \partder{ g_{jl} }{t} \dot{x}^l   \\[1 mm]
           &~& \Ds +  g^{ij} 
            \left[ { \partder{A_0}{x^j} - \partder{A_j}{t} } \right]
             -  g^{ij} 
            \left[ { \partder{A_l}{x^j} - \partder{A_j}{x^l} } \right] 
             \dot{x}^l \deq 0 ~~,
\Ea
}\right.
\nome{EulerLagZeroU}
\Ee 


\renewcommand{\theequation}{B.\arabic{equation}}\setcounter{equation}{0}
\section*{Appendix B: Quaternary constraints of the eleven
fields theory.}

The quaternary constraint of the eleven fields theory are given by:
\begin{eqnarray}
 \xi_\Theta ({\BMz},t) &\equiv&
 \dot{\psi_{\Theta}}({\BMz},t) = \{ \psi_{\Theta} ({\BMz},t) , H_d \} 
\nome{VincoloQuaUa} \\[3 mm]
&=& F^{rs} \left[6\tilde{A}_{,r}\Theta_{,s}
                +3\Theta\nabla_r\nabla_s\tilde{A}
           \right]
   +3\Theta L^A
\nonumber \\[3 mm]
& &\Ds  
  +3\pi^{rs} \Theta 
         \nabla_r \nabla_r \tilde{\lambda}^A
  +6\pi^{rs} \Theta_{,r}\tilde{\lambda}^A_{,s}
  +4\nabla_l\pi^{kl} \Theta\tilde{\lambda}^A_{,k}
\nonumber \\[3 mm]
& &\Ds - \partder{}{z^k} \left[ \Theta(\BMz) p_l g^{lk}(\BMz)  
                           \delta^3\![{\BMz}-{\BMx}(t)]
                    \right] \tilde{\lambda}^A
\nonumber \\[3 mm]
& &\Ds
  +3\pi^{rs}\nabla_r \nabla_s \tilde{A}
            \tilde{\lambda}^\Theta
  +6 \pi^{rs} \tilde{A}_{,r}\tilde{\lambda}^\Theta_{,s}
\nonumber \\[3 mm]
& &\Ds  - 2\nabla_l\pi^{kl} \tilde{A}^2
      \partder{}{z^k}\left[\frac{\tilde{\lambda}^\Theta}{\tilde{A}}
                         \right] 
      +\partder{}{z^k} 
      \left[
      \frac{1}{m} g^{lk} p_l \Hc_M(\BMz) \delta^3\![{\BMz}-{\BMx}(t)] 
      \right] \tilde{\lambda}^\Theta
\nonumber \\[3 mm]
& &\Ds +\frac{2}{m} g^{lk} p_l \Hc_M(\BMz) \delta^3\![{\BMz}-{\BMx}(t)] 
        \tilde{\lambda}^\Theta_{,k}   
\simeq 0 ~~,
\nonumber \\[3 mm]
 \xi_A ({\bf z},t) &\equiv&
 \dot{{\psi}_A}({\BMz},t) = \{ {\psi}_A ({\BMz},t) , H_d \} 
 \nome{VincoloQuaUb} \\[3 mm]
&=& \Ds F^{rs} \left[ 3 \Theta (R_{rs}-\frac{1}{4}g_{rs}R)
                 -6 \frac{\Theta^2}{\tilde{A}^2}
                    \tilde{A}_{,r}\tilde{A}_{,s}
                 -3\Theta\nabla_r\nabla_s\Theta
           \right]
\nonumber \\[3 mm]
& &\Ds + F^{rs}~ 
    \frac{16\pi G\Theta^2}{m\sqrt{g}\tilde{A}^2}
    [p_r p_s - \frac{1}{2} g_{rs}g^{ij}p_i p_j ]
    \delta^3\![{\BMz}-\BMx(t)] 
\nonumber \\[3 mm]
& &\Ds  +3 G_{rs} \Theta
    \left[ \pi^{rs} -\frac{1}{4} \pi g^{rs} \right]
   -3\Theta~ L^\Theta
\nonumber \\[3 mm]
& &\Ds
   +\frac{16\pi G \Theta^2}{\tilde{A}}
    \frac{3}{2} R_{ij}~ 
    \left[ - g^{ij} \Hc_E   
           + \frac{1}{\sqrt{g}}
             (\pi \pi^{ij} - \frac{1}{2} g^{ij}\pi^2)
    \right]   
\nonumber \\[3 mm]
& &\Ds
   -2 \frac{\Theta^2}{\tilde{A}^2}
   \left( 6 \pi^{kl} \tilde{A}_{,l} 
         + \nabla_l\pi^{kl} \tilde{A}
   \right)
   \partder{}{z^k} \left[\frac{\tilde{\lambda}^A}{\tilde{A}}
                   \right]
\nonumber \\[3 mm]
& &\Ds   
   -\frac{32\pi G \Theta^2}{m\sqrt{g}\tilde{A}^3}
    \left[ \pi^{rs} p_r p_s - \frac{1}{2} \pi g^{rs}p_r p_s
    \right] \delta^3\![{\BMz}-\BMx(t)] 
    \tilde{\lambda}^{A}
\nonumber \\[3 mm]
& &\Ds   
   -\frac{(16\pi G)^2 \Theta^3}{m\sqrt{g}\tilde{A}^4}
    \left[ -\frac{1}{\sqrt{g}} 
           \left( \pi \pi^{rs} 
                  +\frac{1}{2}\pi^2 g^{rs}
           \right)
           +\Hc_E g^{rs} 
    \right] p_r p_s \delta^3\![{\BMz}-\BMx(t)] 
    \tilde{\lambda}^{A}
\nonumber \\[3 mm]
& &\Ds   
  +\frac{2}{\tilde{A}^2(\BMz)} m \delta^3\![{\BMz}-\BMx(t)] 
    (\tilde{\lambda}^A)^2 
\nonumber \\[3 mm]
& &\Ds   
  +\frac{2}{\tilde{A}^2(\BMz)} 
    \Hc_M(\BMz) \delta^3\![{\BMz}-\BMx(t)] 
    A_i(\BMz) g^{ij}(\BMz)
    \tilde{\lambda}^A_{,j}
\nonumber \\[3 mm]
& & \Ds
   -6\pi^{rs}\frac{\Theta}{\tilde{A}^2}
     \tilde{A}_{,r}\tilde{A}_{,s}
     \tilde{\lambda}^\Theta
   +3\pi^{rs}(\BMz) \Theta(\BMz)
     \nabla_r\nabla_s \tilde{\lambda}^\Theta
\nonumber \\[3 mm]
& & \Ds
   +\frac{16\pi G \Theta}{m\sqrt{g}\tilde{A}^2}
    \left[ \pi^{rs} p_r p_s - \frac{1}{2} \pi g^{rs}p_r p_s
    \right] \delta^3\![{\BMz}-\BMx(t)] 
    \tilde{\lambda}^\Theta
   +\psi_\Theta \frac{\tilde{\lambda}^\Theta}{\Theta}
\nonumber \\[3 mm]
& &\Ds
  +2 \Theta(\BMz) \nabla_l\pi^{kl}(\BMz)
     \left[ 2 \frac{\tilde{A}_{,k}(\BMz)}{\tilde{A}(\BMz)}
            \tilde{\lambda}^\Theta
           -\tilde{\lambda}^\Theta_{,k}
     \right] 
\nonumber \\[3 mm]
& &\Ds
  - \frac{2}{\tilde{A}^2} \Hc_M
    \delta^3\![{\BMz}-\BMx(t)]
    \tilde{\lambda}^A\tilde{\lambda}^\Theta
\simeq 0 ~~,
\nonumber 
\end{eqnarray}
where $\pi = g_{ij} \pi ^{ij}$, 
$\tilde{\lambda}^A \equiv {\lambda}^A - A_i g^{ij} \tilde{A}_{,j}$,
$\tilde{\lambda}^\Theta \equiv {\lambda}^A - A_i g^{ij} \Theta_{,j}$
and we have defined:
\begin{eqnarray}
F^{rs}(\BMz,t)   &\equiv& \Ds 
       \int d^3\primato{z} ~
             \{\pi^{rs}(\BMz,t) 
               , -\Theta(\primato{\BMz},t)\chi_\Theta(\primato{\BMz},t)
             \}
\nonumber \\[3 mm]
         &=& \Ds \frac{\sqrt{g}}{16\pi G} \left[
                   \Theta \tilde{A}^2 [ R^{rs} - \frac{1}{2} g^{rs} R ]
                   - [ g^{ri}g^{sj} - g^{rs}g^{ij} ]
                     \nabla_i \nabla_j [\Theta\tilde{A}^2]
                   \right] \nonumber \\[2 mm]
         & &\Ds + \frac{\Theta}{2m} g^{ri} g^{sj} p_i p_j
                                \delta^3\! [{\BMz}-{\BMx}(T)]  
\nonumber \\[3 mm]
G_{rs} (\BMz,t)       &\equiv& 
       \int d^3\primato{z}~ 
             \{R_{rs}(\BMz,t) 
               , -\Theta(\primato{\BMz},t)\chi_\Theta(\primato{\BMz},t)
             \}
\nonumber \\[3 mm]
& = & \Ds 
            (\hat{\delta}_{rs}^{im} g^{jn}
            +\hat{\delta}_{rs}^{in} g^{jm}
            -\hat{\delta}_{rs}^{ij} g^{mn}
            -\hat{\delta}_{rs}^{mn} g^{ij})
      \nabla_i \nabla_j \left[
           \frac{16 \pi G\Theta}{\sqrt{g}\tilde{A}}
           (\pi_{mn} - \frac{1}{2} g_{mn} \pi )
           \right]  
\nonumber \\[3 mm]
& & \Ds -\frac{1}{4} g^{km}  
            (\hat{\delta}_{rl}^{ij} R^{l}_{~mks}
            +\hat{\delta}_{sl}^{ij} R^{l}_{~mkr}
            -\hat{\delta}_{kl}^{ij} R^{l}_{~rsm}
            -\hat{\delta}_{kl}^{ij} R^{l}_{~srm}
            +4 \hat{\delta}_{kl}^{ij} R^{l}_{~rms} )
\nonumber \\[3 mm]
& & \Ds \qquad\qquad
           \left[
           \frac{16 \pi G\Theta}{\sqrt{g}\tilde{A}}
           (\pi_{ij} - \frac{1}{2} g_{ij} \pi )
           \right]  
\nonumber \\[3 mm]
L^A (\BMz,t)     &\equiv& \Ds
       \int d^3\primato{z} ~\pi^{rs}(\BMz,t)
             \{\nabla_r\nabla_s \tilde{A}(\BMz,t) 
               , -\Theta(\primato{\BMz},t)\chi_\Theta(\primato{\BMz},t)
             \}
\nonumber \\[3 mm]
 &=&\Ds \frac{16\pi G\Theta\tilde{A}_{,k}}{\tilde{A}}
        \left[\frac{\Theta_{,l}}{\Theta}
             -\frac{\tilde{A}_{,k}}{\tilde{A}}
        \right]
        \left[\frac{2}{\sqrt{g}} (\pi^{ki}\pi^{lj}g_{ij}
                                  -\frac{1}{2} \pi^{kl} \pi )
             -g^{lk} \Hc_E 
       \right]
\nonumber \\[3 mm]
 & &\Ds+\frac{16\pi G\Theta\tilde{A}_{,k}}{\tilde{A}}
        \left[ 2 g_{lj} \pi^{ij}
              \nabla_i [ \pi^{kl}-\frac{1}{2}g^{kl}\pi ]
              - [ \pi_{ab} -\frac{1}{2}g_{ab}\pi ]
                g^{ki} \nabla_i \pi^{ab}
        \right]
\nonumber \\[3 mm]
L^\Theta (\BMz,t) &\equiv& \Ds
       \int d^3\primato{z} ~\pi^{rs}(\BMz,t)
             \{\nabla_r\nabla_s \Theta(\BMz,t) 
               , -\Theta(\primato{\BMz},t)\chi_\Theta(\primato{\BMz},t)
             \}
\nonumber \\[3 mm]
 &=&\Ds \frac{16\pi G\Theta\Theta_{,k}}{\tilde{A}}
        \left[\frac{\Theta_{,l}}{\Theta}
             -\frac{\tilde{A}_{,k}}{\tilde{A}}
        \right]
        \left[\frac{2}{\sqrt{g}} (\pi^{ki}\pi^{lj}g_{ij}
                                  -\frac{1}{2} \pi^{kl} \pi )
             -g^{lk} \Hc_E 
       \right]
\nonumber \\[3 mm]
 & &\Ds+\frac{16\pi G\Theta\Theta_{,k}}{\tilde{A}}
        \left[ 2 g_{lj} \pi^{ij}
              \nabla_i [ \pi^{kl}-\frac{1}{2}g^{kl}\pi ]
              - [ \pi_{ab} -\frac{1}{2}g_{ab}\pi ]
                g^{ki} \nabla_i \pi^{ab}
        \right]
\nonumber 
\end{eqnarray}
and $\hat{\delta}^{rs}_{ij}$ is the symmetrized expression 
$\frac{1}{2} (\delta^r_i \delta^s_j + \delta^s_i \delta^r_j)$.


\renewcommand{\theequation}{C.\arabic{equation}}\setcounter{equation}{0}
\section*{Appendix C: Constraints algebra of the eleven-fields theory}

We summarize here the relevant part of the constraints algebra of the
eleven fields theory:
\begin{eqnarray*}
  \{ \chi_A({\BMz},t)       , \phi_i(\primato{{\BMz}},t)      \} 
&=&\Ds  
  + \tilde{A}_{,i}(\primato{{\BMz}},t) 
    \{ \chi_A({\BMz},t)       , \pi_A(\primato{{\BMz}},t)       \} 
\\[4 mm]
& &\Ds 
  + \Theta_{,i}    (\primato{{\BMz}},t) 
    \{ \chi_A({\BMz},t)      , \pi_\Theta(\primato{{\BMz}},t)  \} 
\\[4 mm]
& &\Ds 
  - \chi_A(\primato{{\BMz}},t) 
    \partial_i \delta^3\![{\BMz}-\primato{{\BMz}}] 
\\[4 mm] 
  \{ \chi_\Theta({\BMz},t)  , \phi_i(\primato{{\BMz}},t)      \} 
&=&\Ds  
  + \tilde{A}_{,i}(\primato{{\BMz}},t) 
    \{ \chi_\Theta({\BMz},t)  , \pi_A(\primato{{\BMz}},t)       \} 
\\[4 mm]
& &\Ds 
  + \Theta_{,i}   (\primato{{\BMz}},t) 
    \{ \chi_\Theta({\BMz},t)  , \pi_\Theta(\primato{{\BMz}},t)  \} 
\\[3 mm] 
& & \Ds
   - \chi_\Theta(\primato{{\BMz}},t) 
     \partial_i \delta^3\![{\BMz}-\primato{{\BMz}}] 
\\[4 mm]
  \{ \chi_A({\BMz},t)       , \pi_A(\primato{{\BMz}},t)       \} 
&=&\Ds   
  \frac{2}{\tilde{A}^2} 
  \left[{\chi_\Theta 
     +\left( \frac{1}{2m}g^{ij}p_i p_j - m \tilde{A} \right) 
      \delta^3\! [{\BMz}-{\BMx}(T)]
       }\right] 
     \delta^3\![{\BMz}-\primato{{\BMz}}]  
\\[3 mm]
  \{ \chi_\Theta({\BMz},t)  , \pi_A(\primato{{\BMz}},t)       \} 
&=&\Ds   
  \frac{1}{\Theta} \chi_A  \delta^3\![{\BMz}-\primato{{\BMz}}]  
\\[3 mm]
  \{ \chi_A({\BMz},t)       , \pi_\Theta(\primato{{\BMz}},t)  \} 
&=&\Ds   
  \frac{1}{\Theta} \chi_A  \delta^3\![{\BMz}-\primato{{\BMz}}]  
\\[4 mm]
  \{ \chi_\Theta({\BMz},t)  , \pi_\Theta(\primato{{\BMz}},t)   \} 
&=&\Ds     0  
\\[4 mm]
  \{ \psi_A({\BMz},t)       , \phi_i(\primato{{\BMz}},t)      \} 
&=&\Ds  
  + \tilde{A}_{,i}(\primato{{\BMz}},t) 
    \{ \psi_A({\BMz},t)       , \pi_A(\primato{{\BMz}},t)       \} 
\\[4 mm]
& &\Ds 
  + \Theta_{,i}    (\primato{{\BMz}},t) 
    \{ \psi_A({\BMz},t)      , \pi_\Theta(\primato{{\BMz}},t)  \} 
\\[4 mm]
& &\Ds 
  - \psi_A(\primato{{\BMz}},t) 
    \partial_i \delta^3\![{\BMz}-\primato{{\BMz}}] 
\\[4 mm] 
  \{ \psi_\Theta({\BMz},t)  , \phi_i(\primato{{\BMz}},t)      \} 
&=&\Ds  
  + \tilde{A}_{,i}(\primato{{\BMz}},t) 
    \{ \psi_\Theta({\BMz},t)  , \pi_A(\primato{{\BMz}},t)       \} 
\\[4 mm]
& &\Ds 
  + \Theta_{,i}   (\primato{{\BMz}},t) 
    \{ \psi_\Theta({\BMz},t)  , \pi_\Theta(\primato{{\BMz}},t)  \} 
\\[3 mm] 
& & \Ds
   - \psi_\Theta(\primato{{\BMz}},t) 
     \partial_i \delta^3\![{\BMz}-\primato{{\BMz}}] 
\\[4 mm]
  \{ \chi_A({\BMz},t)       , \chi_A(\primato{{\BMz}},t)      \} 
&=&\Ds
   -2 \frac{\Theta^2(\BMz)}{\tilde{A}(\BMz)}                
   \left[\nabla_l\pi^{kl} (\BMz)
         -3\tilde{A}_{,l} (\BMz) \pi^{kl}(\BMz)
   \right] \partial_k \delta^3\![{\BMz}-\primato{{\BMz}}]   
\\[4 mm]
& &\Ds 
   -2 \frac{\Theta^2(\primato{\BMz})}{\tilde{A}(\primato{\BMz})}
   \left[\nabla_l\pi^{kl} (\primato{\BMz})
         -3\tilde{A}_{,l} (\primato{\BMz}) \pi^{kl}(\primato{\BMz})
   \right] \partial_k \delta^3\![{\BMz}-\primato{{\BMz}}]   
\\[4 mm]
  \{ \chi_A({\BMz},t)       , \chi_\Theta(\primato{{\BMz}},t) \} 
&=& \Ds 
   - 3\Theta \left[ \pi^{ij} R_{ij} - \frac{1}{4} \pi R \right]
     \delta^3\![{\BMz}-\primato{{\BMz}}]  
   + 6 \pi^{ij} \frac{\Theta}{\tilde{A}^2} \tilde{A}_{,i}\tilde{A}_{,j} 
     \delta^3\![{\BMz}-\primato{{\BMz}}]   
\\[4 mm]
& &\Ds  
   + 3 \Theta({\BMz},t)  \pi^{ij}({\BMz},t)  
   \left[\partial_i\partial_j  \delta^3\![{\BMz}-\primato{{\BMz}}]   
      -\Gamma^k_{ij}({\BMz},t) 
       \partial_k \delta^3\![{\BMz}-\primato{{\BMz}}]   
   \right]   
\\[4 mm]
& &\Ds  
    -\frac{16\pi G}{m}\frac{\Theta}{\sqrt{g}\tilde{A}^2}
     \left[ \pi^{ij} -\frac{1}{2} g^{ij} \pi \right] p_i p_j 
     \delta^3\![{{\BMz}}-{\BMx}(t)]
     \delta^3\![{\BMz}-\primato{{\BMz}}]   
\\[4 mm]
& &\Ds  
   +\left[ 2 \Theta(\BMz) \nabla_l\pi^{kl} (\BMz)
   \right] \partial_k \delta^3\![{\BMz}-\primato{{\BMz}}]   
   - 2 \Theta(\BMz) \nabla_l\pi^{kl} (\BMz)
     \frac{\tilde{A}_{,l}(\BMz)}{\tilde{A}(\BMz)}
     \delta^3\![{\BMz}-\primato{{\BMz}}]   
\\[4 mm] 
& & + \Ds
   \Theta(\primato{{\BMz}})   
   \left[ 2 \nabla_l\pi^{kl} (\primato{{\BMz}})
         + p_l g^{lk}(\primato{{\BMz}}) 
           \delta^3\![\primato{{\BMz}}-{\BMx}(t)]
   \right] \partial_k \delta^3\![{\BMz}-\primato{{\BMz}}]   
\\[4 mm] 
& &\Ds 
    - \Theta_{,i}({\BMz},t) 
      \left[
        2 \pi^{ij}_{|j}(\BMz,t)
      + p_j g^{ij}({\BMz},t) \delta^3\![{{\BMz}}-{\BMx}(t)]
      \right] \delta^3\![{\BMz}-\primato{{\BMz}}] 
\\[4 mm] 
  \{ \chi_\Theta({\BMz},t)  , \chi_\Theta(\primato{{\BMz}},t) \} 
&=&\Ds 
   ~\left[  \tilde{A}(\BMz) \nabla_l\pi^{kl} (\BMz)
         -3\tilde{A}_{,l} (\BMz) \pi^{kl}(\BMz)
   \right] \partial_k \delta^3\![{\BMz}-\primato{{\BMz}}]   
\\[4 mm]
 & &+\Ds 
   \left[  \tilde{A}(\primato{\BMz}) \nabla_l\pi^{kl} (\primato{\BMz})
         -3\tilde{A}_{,l} (\primato{\BMz}) \pi^{kl}(\primato{\BMz})
   \right] \partial_k \delta^3\![{\BMz}-\primato{{\BMz}}]   
\\[4 mm]
 & &-\Ds
   \left[ 
         \frac{p_l g^{rk}(\BMz,t)}{m}
          \Hc_M ({{\BMz}},t)
         +\frac{p_l g^{rk}(\primato{{\BMz}},t)}{m} 
          \Hc_M (\primato{{\BMz}},t)
   \right] \partial_k \delta^3\![{\BMz}-\primato{{\BMz}}]   
\\[4 mm]
  \{ \psi_A({\BMz},t)       , \pi_A(\primato{{\BMz}},t)       \} 
&=&\Ds   
   -12 \pi^{kl} \frac{\Theta^2}{\tilde{A}^2} 
      \tilde{A}_{,k} \partial_l \delta^3\![{\BMz}-\primato{{\BMz}}] 
\\[3 mm]
& &\Ds   
   +12 \pi^{kl} \frac{\Theta^2}{\tilde{A}^3} 
      \tilde{A}_{,k} \tilde{A}_{,l} \delta^3\![{\BMz}-\primato{{\BMz}}] 
\\[3 mm]
& &\Ds   
   -2 \frac{\Theta^2}{\tilde{A}} \nabla_l\pi^{kl} 
      \partial_k \delta^3\![{\BMz}-\primato{{\BMz}}] 
\\[3 mm]
& &\Ds   
   +2 \frac{\Theta^2}{\tilde{A}^2} \nabla_l\pi^{kl} 
      \tilde{A}_{,k} \delta^3\![{\BMz}-\primato{{\BMz}}] 
\\[3 mm]
& &\Ds   
   -\frac{32\pi G \Theta^2}{m\sqrt{g}\tilde{A}^3}
    \left[ \pi^{rs} p_r p_s - \frac{1}{2} \pi g^{rs}p_r p_s
    \right] \delta^3\![{\BMz}-\BMx(t)] 
    \delta^3\![{\BMz}-\primato{{\BMz}}] 
\\[3 mm]
& &\Ds   
  +\frac{2}{\tilde{A}^2(\BMz)} m \delta^3\![{\BMz}-\BMx(t)] 
    \left[ \lambda^A(\BMz) - A_i g^{ij}(\BMz) \tilde{A}_{,j}(\BMz)
    \right]
    \delta^3\![{\BMz}-\primato{{\BMz}}] 
\\[3 mm]
& &\Ds   
  +\frac{2}{\tilde{A}^2(\BMz)} 
    \Hc_M(\BMz) \delta^3\![{\BMz}-\BMx(t)] 
    A_i(\BMz) g^{ij}(\BMz)
    \partial_j \delta^3\![{\BMz}-\primato{{\BMz}}] 
\\[3 mm]
  \{ \psi_A({\BMz},t)       , \pi_\Theta(\primato{{\BMz}},t)  \} 
&=& \Ds
    \left\{ 
           3 \pi^{rs} \left[ R_{rs} - \frac{1}{4} g_{rs} R \right]
          -3\pi^{rs} \nabla_r\nabla_s \Theta 
    \right\} 
    \delta^3\![{\BMz}-\primato{{\BMz}}] 
\\[3 mm]
& &\Ds
   -12\pi^{rs}\frac{\Theta}{\tilde{A}^2}
     \tilde{A}_{,r}\tilde{A}_{,s}
     \delta^3\![{\BMz}-\primato{{\BMz}}]    
\\[3 mm]
& &\Ds
   +3\pi^{rs}(\BMz) \Theta(\BMz)
     \partial_r\partial_s
     \delta^3\![{\BMz}-\primato{{\BMz}}]    
\\[3 mm]
& &\Ds
   -3\pi^{rs}(\BMz) \Theta(\BMz)
     \Gamma^k_{rs}(\BMz)
     \partial_k\delta^3\![{\BMz}-\primato{{\BMz}}]    
\\[3 mm]
& &\Ds   
   -\frac{32\pi G \Theta}{m\sqrt{g}\tilde{A}^2}
    \left[ \pi^{rs} p_r p_s - \frac{1}{2} \pi g^{rs}p_r p_s
    \right] \delta^3\![{\BMz}-\BMx(t)] 
    \delta^3\![{\BMz}-\primato{{\BMz}}] 
\\[3 mm]
& &\Ds
  +2 \Theta(\BMz) \nabla_l\pi^{kl}(\BMz)
     \left[ 2 \frac{\tilde{A}_{,k}(\BMz)}{\tilde{A}(\BMz)}
           -\frac{\Theta_{,k}(\BMz)}{\Theta(\BMz)}
     \right] 
    \delta^3\![{\BMz}-\primato{{\BMz}}] 
\\[3 mm]
& &\Ds
  -2 \Theta(\BMz) \nabla_l\pi^{kl}(\BMz)
    \partial_k \delta^3\![{\BMz}-\primato{{\BMz}}] 
\\[3 mm]
  \{ \psi_\Theta({\BMz},t)  , \pi_A(\primato{{\BMz}},t)       \} 
&=&\Ds  
   3\pi^{rs}(\BMz) \Theta(\BMz) \left[
         \partial_r \partial_s \delta^3\![{\BMz}-\primato{{\BMz}}]
        -\Gamma^k_{rs}\partial_k \delta^3\![{\BMz}-\primato{{\BMz}}]
        \right]
\\[3 mm]
& &\Ds +6 \pi^{rs}(\BMz) \partial_r \Theta(\BMz)
                    \partial_s \delta^3\![{\BMz}-\primato{{\BMz}}]
\\[3 mm]
& &\Ds +\Theta(\BMz) \left[ 2\nabla_l\pi^{kl}(\BMz) 
                      -p_l g^{lk}(\BMz)  \delta^3\![{\BMz}-{\BMx}(t)]
               \right]\partial_k\delta^3\![{\BMz}-\primato{{\BMz}}]
\\[3 mm]
& &\Ds - \partder{}{z^k} \left[ \Theta(\BMz) p_l g^{lk}(\BMz)  
                           \delta^3\![{\BMz}-{\BMx}(t)]
                    \right] \delta^3\![{\BMz}-\primato{{\BMz}}]
\\[3 mm]
& &\Ds -  \left[ 2\nabla_l\pi^{kl}(\BMz) 
           +p_l g^{lk}(\BMz)  \delta^3\![{\BMz}-{\BMx}(t)]
               \right]
     \partial_k\Theta(\BMz) \delta^3\![{\BMz}-\primato{{\BMz}}]
\\[3 mm]
  \{ \psi_\Theta({\BMz},t)  , \pi_\Theta(\primato{{\BMz}},t)  \} 
&=&\Ds   
   3\pi^{rs}(\BMz) \nabla_r \nabla_s \tilde{A}(\BMz) 
                   \delta^3\![{\BMz}-\primato{{\BMz}}]
  +6 \pi^{rs}(\BMz) \partial_r \tilde{A}(\BMz)
                    \partial_s \delta^3\![{\BMz}-\primato{{\BMz}}]
\\[3 mm]
& &\Ds + 2\nabla_l\pi^{kl}(\BMz) \partial_k \tilde{A}(\BMz)
         \delta^3\![{\BMz}-{\BMx}(t)]
       - 2\nabla_l\pi^{kl}(\BMz) \tilde{A}(\BMz)
         \partial_k \delta^3\![{\BMz}-{\BMx}(t)]
\\[3 mm]
& &\Ds + \partder{}{z^k} \left[
      \frac{1}{m} g^{lk} p_l \Hc_M(\BMz) \delta^3\![{\BMz}-{\BMx}(t)] 
      \right] \delta^3\![{\BMz}-\primato{{\BMz}}]
\\[3 mm]
& &\Ds +\frac{2}{m} g^{lk} p_l \Hc_M(\BMz) \delta^3\![{\BMz}-{\BMx}(t)] 
        \partial_k \delta^3\![{\BMz}-\primato{{\BMz}}]
\\[3 mm]
& &\Ds -  \left[ 2\nabla_l\pi^{kl}(\BMz) 
           +p_l g^{lk}(\BMz)  \delta^3\![{\BMz}-{\BMx}(t)]
               \right]
     \partial_k\Theta(\BMz) \delta^3\![{\BMz}-\primato{{\BMz}}]
\\[3 mm]
\end{eqnarray*}


\def\vol#1{{{\bf #1}}}


\end{document}